\documentclass[twocolumn]{aastex631}

\newcommand{\radex}{\texttt{SpectralRadex}}
\newcommand{\astrodendro}{\texttt{astrodendro}}
\newcommand{\spectuner}{\texttt{spectuner}}

\newcommand{\HII}{HC/UC\,H\,{\sc ii}}
\newcommand{\CO}{$^{12}$CO\,(2--1)}
\newcommand{\kms}{km\,s$\rm ^{-1}$}

\newcommand{\degree}{$^{\circ}$}
\newcommand{\rms}{$\sigma$}

\def \ch3cn{CH$_3$CN}
\def \12CO{$^{12}$CO}

\usepackage{newunicodechar}
\usepackage{amsmath}
\usepackage{color}

\DeclareUnicodeCharacter {2212} {-}
\DeclareUnicodeCharacter {02BC} {-}

\usepackage{txfonts}

\usepackage{graphicx}	
\usepackage{amsmath}	
\usepackage{color}
\usepackage{float}
\usepackage{soul}
\usepackage{multirow}

\usepackage{booktabs}
\usepackage{longtable}
\usepackage{xcolor}

\usepackage{xparse}
\usepackage{xcolor}
\usepackage[normalem]{ulem}
\NewDocumentCommand{\fengwei}{ m g }{%
    \leavevmode
    \IfNoValueTF{#2}
        {\textcolor{teal}{[Fengwei: #1]}}
        {{\color{red}\sout{#1}} \textcolor{teal}{[Fengwei: #2]}}
}

\begin{document}

\title{The ALMA-QUARKS survey: Hot Molecular Cores are a long-standing phenomenon in the evolution of massive protostars}


\author[0009-0000-5764-8527]{Dezhao Meng}\thanks{E-mail:mengdezhao@xao.ac.cn}
\affiliation{Xinjiang Astronomical Observatory, Chinese Academy of Sciences, Urumqi 830011, People’s Republic of China}
\affiliation{University of Chinese Academy of Sciences, Beijing 100080,
People’s Republic of China}
\affiliation{Shanghai Astronomical Observatory, Chinese Academy 
of Sciences, Shanghai 200030, People’s Republic of China}

\author[0000-0002-5286-2564]{Tie Liu}\thanks{E-mail:liutie@shao.ac.cn}
\affiliation{Shanghai Astronomical Observatory, Chinese Academy 
of Sciences, Shanghai 200030, People’s Republic of China}
\affiliation{State Key Laboratory of Radio Astronomy and Technology, Beijing 100101, People’s Republic of China}

\author[0000-0003-4910-1390]{Jarken Esimbek}\thanks{E-mail:jarken@xao.ac.cn}
\affiliation{Xinjiang Astronomical Observatory, Chinese Academy of Sciences, Urumqi 830011, People’s Republic of China}
\affiliation{State Key Laboratory of Radio Astronomy and Technology, Beijing 100101, People’s Republic of China}
\affiliation{Xinjiang Key Laboratory of Radio Astrophysics, Urumqi 830011, People’s Republic of China}

\author[0000-0003-2302-0613]{Sheng-Li Qin}
\affiliation{School of Physics and Astronomy, Yunnan University, Kunming 650091, People’s Republic of China}

\author{Guido Garay}
\affiliation{Departamento de Astronom\'ia, Universidad de Chile, Las Condes, 7591245 Santiago, Chile}
\affiliation{Chinese Academy of Sciences South America Center for Astronomy, National Astronomical Observatories, CAS, Beijing 100101, People’s Republic of China}

\author[0000-0002-6622-8396]{Paul F. Goldsmith}
\affiliation{Jet Propulsion Laboratory, California Institute of Technology, 4800 Oak Grove Drive, Pasadena, CA 91109, USA}

\author[0000-0003-0356-818X]{Jianjun Zhou}
\affiliation{Xinjiang Astronomical Observatory, Chinese Academy of Sciences, Urumqi 830011, People’s Republic of China}
\affiliation{State Key Laboratory of Radio Astronomy and Technology, Beijing 100101, People’s Republic of China}
\affiliation{Xinjiang Key Laboratory of Radio Astrophysics, Urumqi 830011, People’s Republic of China}

\author[0000-0002-4154-4309]{Xindi Tang}
\affiliation{Xinjiang Astronomical Observatory, Chinese Academy of Sciences, Urumqi 830011, People’s Republic of China}
\affiliation{State Key Laboratory of Radio Astronomy and Technology, Beijing 100101, People’s Republic of China}
\affiliation{Xinjiang Key Laboratory of Radio Astrophysics, Urumqi 830011, People’s Republic of China}

\author[0000-0001-9822-7817]{Wenyu Jiao}
\affiliation{Shanghai Astronomical Observatory, Chinese Academy of Sciences, Shanghai 200030, People’s Republic of China}

\author[0000-0001-7817-1975]{Yan-Kun Zhang}
\affiliation{Shanghai Astronomical Observatory, Chinese Academy of Sciences, Shanghai 200030, People’s Republic of China}

\author[0000-0001-5950-1932]{Fengwei Xu}
\affiliation{Kavli Institute for Astronomy and Astrophysics, Peking University, 5 Yiheyuan Road, Haidian District, Beijing 100871, People’s Republic of China}

\author[0000-0002-9836-0279]{Siju Zhang}
\affiliation{Departamento de Astronom\'ia, Universidad de Chile, Las Condes, 7591245 Santiago, Chile}

\author[0000-0001-5917-5751]{Anandmayee Tej}
\affiliation{Indian Institute of Space Science and Technology, Thiruvananthapuram 695 547, India}

\author[0000-0002-9574-8454]{Leonardo Bronfman}
\affiliation{Departamento de Astronom\'ia, Universidad de Chile, Las Condes, 7591245 Santiago, Chile}

\author[0000-0003-4546-2623]{Aiyuan Yang}
\affiliation{National Astronomical Observatories, Chinese Academy of Sciences, A20 Datun Road, Chaoyang District, Beijing, 100101, People’s Republic of China}

\author[0000-0002-8697-9808]{Sami Dib}
\affiliation{Max Planck Institute for Astronomy, K\"{o}nigstuhl 17, 69117, Heidelberg, Germany}

\author[0000-0001-7151-0882]{Swagat R. Das}
\affiliation{Departamento de Astronom\'ia, Universidad de Chile, Las Condes, 7591245 Santiago, Chile}

\author[0000-0001-7866-2686]{Jihye Hwang}
\affil{Institute for Advanced Study, Kyushu University, Japan}
\affil{Department of Earth and Planetary Sciences, Faculty of Science, Kyushu University, Nishi-ku, Fukuoka 819-0395, Japan}

\author{Archana Soam}
\affil{Indian Institute of Astrophysics, II Block, Koramangala, Bengaluru 560034, India}

\author[0000-0002-7716-1094]{Yisheng Qiu}
\affiliation{Research Center for Astronomical computing, Zhejiang Laboratory, Hangzhou 311121, China}

\author[0000-0001-5494-6238]{Dalei Li}
\affiliation{Xinjiang Astronomical Observatory, Chinese Academy of Sciences, Urumqi 830011, People’s Republic of China}
\affiliation{State Key Laboratory of Radio Astronomy and Technology, Beijing 100101, People’s Republic of China}
\affiliation{Xinjiang Key Laboratory of Radio Astrophysics, Urumqi 830011, People’s Republic of China}

\author[0000-0002-8760-8988]{Yuxin He}
\affiliation{Xinjiang Astronomical Observatory, Chinese Academy of Sciences, Urumqi 830011, People’s Republic of China}
\affiliation{State Key Laboratory of Radio Astronomy and Technology, Beijing 100101, People’s Republic of China}
\affiliation{Xinjiang Key Laboratory of Radio Astrophysics, Urumqi 830011, People’s Republic of China}

\author[0000-0003-0933-7112]{Gang Wu}
\affiliation{Xinjiang Astronomical Observatory, Chinese Academy of Sciences, Urumqi 830011, People’s Republic of China}
\affiliation{State Key Laboratory of Radio Astronomy and Technology, Beijing 100101, People’s Republic of China}
\affiliation{Xinjiang Key Laboratory of Radio Astrophysics, Urumqi 830011, People’s Republic of China}

\author[0000-0001-6725-0483]{Lokesh Dewangan}
\affiliation{Physical Research Laboratory, Navrangpura, Ahmedabad 380009, India}

\author[0000-0002-9875-7436]{James O. Chibueze}
\affiliation{Department of Mathematical Sciences, University of South Africa, Cnr Christian de Wet Rd and Pioneer Avenue, Florida Park, 1709, Roodepoort, South Africa}
\affiliation{Department of Physics and Astronomy, Faculty of Physical Sciences, University of Nigeria, Carver Building, 1 University Road, Nsukka 410001, Nigeria}

\author[0000-0002-8586-6721]{Pablo Garc{\'i}a}
\affiliation{Chinese Academy of Sciences South America Center for Astronomy, National Astronomical Observatories, CAS, Beijing 100101, People’s Republic of China}
\affiliation{Instituto de Astronom\'ia, Universidad Cat\'olica del Norte, Av. Angamos 0610, Antofagasta, Chile}

\author{Prasanta Gorai}
\affiliation{Rosseland Centre for Solar Physics, University of Oslo, PO Box 1029 Blindern, 0315 Oslo, Norway}
\affiliation{Institute of Theoretical Astrophysics, University of Oslo, PO Box 1029 Blindern, 0315 Oslo, Norway}

\author[0000-0001-8812-8460]{Naval Kishor Bhadari}
\affiliation{Kavli Institute for Astronomy and Astrophysics, Peking University, 5 Yiheyuan Road, Haidian District, Beijing 100871, People’s Republic of China}

\author[0000-0002-1086-7922]{Yong Zhang}
\affiliation{School of Physics and Astronomy, Sun Yat-sen University, 2 Daxue Road, Tangjia, Zhuhai, Guangdong Province, People’s Republic of China}

\author[0000-0002-7125-7685]{Patricio Sanhueza}
\affiliation{Department of Astronomy, School of Science, The University of Tokyo, 7-3-1 Hongo, Bunkyo-ku, Tokyo 113-0033, Japan}

\author[0009-0000-8349-7355]{Yongquan Luo}
\affiliation{Xinjiang Astronomical Observatory, Chinese Academy of Sciences, Urumqi 830011, People’s Republic of China}
\affiliation{University of Chinese Academy of Sciences, Beijing 100080,
People’s Republic of China}

\author[0009-0000-9090-9960]{Jia-Hang Zou}
\affiliation{Shanghai Astronomical Observatory, Chinese Academy of Sciences, Shanghai 200030, People’s Republic of China}
\affiliation{School of Physics and Astronomy, Yunnan University, Kunming 650091, People’s Republic of China}

\author{Kee-Tae Kim}
\affiliation{Korea Astronomy and Space Science Institute, 776 Daedeokdaero, Yuseong-gu, Daejeon 34055, Republic of Korea}
\affiliation{University of Science and Technology, Korea (UST), 217 Gajeong-ro, Yuseong-gu, Daejeon 34113, Republic of Korea}

\author{Dongting Yang}
\affiliation{School of Physics and Astronomy, Yunnan University, Kunming 650091, People’s Republic of China}

\author{Lei Zhu}
\affiliation{Chinese Academy of Sciences South America Center for Astronomy, National Astronomical Observatories, CAS, Beijing 100101, People’s Republic of China}

\author[0000-0001-8315-4248]{Xunchuan Liu}
\affiliation{Shanghai Astronomical Observatory, Chinese Academy 
of Sciences, Shanghai 200030, People’s Republic of China}

\author{Macleod Gordon}
\affiliation{Xinjiang Astronomical Observatory, Chinese Academy of Sciences, Urumqi 830011, People’s Republic of China}


\begin{abstract}
We present an analysis of the QUARKS survey sample, focusing on protoclusters where Hot Molecular Cores (HMCs, traced by \ch3cn\,(12--11)) and \HII\ regions (traced by H30$\alpha$/H40$\alpha$) coexist.
\textcolor{black}{
Using the high-resolution, high-sensitivity 1.3~mm data from the QUARKS survey, we identify 125 Hot Molecular Fragments (HMFs), which represent the substructures of HMCs at higher resolution.}
\textcolor{black}{From} line integrated intensity maps of \textcolor{black}{\ch3cn\,(12$_3$--11$_3$)} and H30$\alpha$, we resolve the spatial distribution of HMFs and \HII\ regions. By combining with observations of \CO\ outflows and 1.3 mm continuum, we classify \textcolor{black}{HMFs into four types: HMFs associated with jet-like outflow, with wide-angle outflow, with non-detectable outflow, and shell-like HMFs near \HII\ regions}.
This diversity possibly indicates that the hot core could be polymorphic and long-standing phenomenon in the evolution of massive protostars.
The separation between \textcolor{black}{HMFs} and H30$\alpha$/H40$\alpha$ emission suggests that sequential high-mass star formation within young protoclusters is not likely related to feedback mechanisms.
\end{abstract}
\keywords{Interstellar medium (847) --- Star forming regions (1565) --- Star formation (1569) --- Protostars (1302) }

\section{Introduction}
\label{Introduction}

High-mass stars ($M_{\star} \gtrsim 8 M_{\odot}$) play a crucial role in galactic evolution by shaping galactic structures and serving as primary sources of heavy elements and ultraviolet (UV) radiation in the interstellar medium (ISM) \citep{Zin2007}. However, the formation and evolution of high-mass stars remain poorly understood.
Newly formed high-mass protostars heat their surrounding material with temperatures exceeding 100 K, forming hot cores -- compact, dense regions ($\lesssim$\,0.1\,pc, n$_{\mathrm{H_2}} \gtrsim$ 10$^6~\mathrm{cm}^{-3}$, e.g., \citealt{Gar1999,May1999,Kur2000,Ces2005,QSL2022,LZY2025}), rich in complex organic molecules (COMs, refer to molecules containing carbon and consisting of six or more atoms).
These hot cores are thought to represent an early evolutionary stage of high-mass star formation, being precursors to evolved H\,{\sc ii} regions \citep[e.g.,][]{Beu2007,Purcell2006,Gerner2014,CHoudhury2015,Gerner2015, Miyawaki2021}.
Hot cores provide critical insights into high-mass star formation, as the detected COMs serve as excellent tracers of both the physical conditions and chemical processes occurring in the immediate environments of high-mass protostars \citep{Bel2018,Jor2020,Gie2021,Wil2022,TK2023,JS2025,ST2025}.

Extensive studies of hot cores have been carried out with single dish radio telescopes, but are limited by poor angular resolution \citep{Sch1997,Gib2000,Van2000,Sch2001,Sch2006,Bis2007,Fon2007,Bel2013,Hal2013,Cro2014,Nei2014,Mol2021,NS2021}. Recent advances in millimeter/submillimeter interferometric arrays (e.g., SMA, NOEMA, and ALMA), with their superior spatial resolution and sensitivity, have enabled detailed investigations of the physical and chemical properties of hot cores. However, most interferometric studies of hot cores have focused on either individual sources or small samples \citep{LSY2001,LSY2002,QSL2008,Beu2009,QSL2010,WYF2014,QSL2015,Bog2019,Mot2020,Fue2021,Gie2021,Law2021,Van2021,atoms15,assemble}. The ALMA Three-millimeter Observations of Massive Star-forming regions (ATOMS, Project ID: 2019.1.00685.S; PI: Tie Liu) survey \citep{LT2020} recently targeted 146 active high-mass star-forming regions at moderately high angular resolution ($\sim$\,2\arcsec). Based on the ATOMS survey, \citet{QSL2022} identified 60 hot cores by using C$_2$H$_5$CN, CH$_3$OCHO, and CH$_3$OH lines. Notably, in 24 of these hot cores, the COMs emission distributions were spatially offset from the \textcolor{black}{3 mm} continuum emission, particularly where the \textcolor{black}{3 mm} continuum emission contains contributions from both dust and free-free emission associated with hypercompact/ultracompact H\,{\sc ii} (\HII) regions. 
Nevertheless, the limited resolution of ATOMS prevents definitive discrimination between two possible scenarios for the COMs emission near \HII\ regions: (1) residual molecular gas \textcolor{black}{adjacent to} the ionized region, versus (2) a separate source at a different evolutionary stage \citep{JS2025}.
In the latter case, an important question arises as to whether the hot cores around these \HII\ regions could have been triggered by \HII\ feedback. Radiation pressure and stellar winds from massive stars can compress the interstellar medium and triggering subsequent star formation in dense layers. This so-called “collect-and-collapse” process has been revealed at the borders of several \HII\ regions over the past decades \citep{Deharveng2003,Deharveng2005,Zavagno2006,Zavagno2007,Deharveng2008,Pomares2009,Petriella2010,Brand2011,Liu2012,LHL2015,LHL2016,LT2017,zhou2020}. A high-resolution, statistically significant survey of hot cores around \HII\ regions is therefore crucial for exploring the sequential star formation within young protoclusters.

The Querying Underlying mechanisms of massive star formation with ALMA-Resolved gas Kinematics and Structures (QUARKS, Project ID: 2021.1.00095.S; PIs: Lei Zhu, Guido Garay and Tie Liu) survey \citep{LXC2024, quarks2} has observed 139 protoclusters (selected from the ATOMS sample) at 1.3 mm, achieving higher angular resolution and more comprehensive molecular line coverage than the ATOMS survey. Thus, we utilize the QUARKS data to resolve the spatial distribution of COMs in the vicinity of \HII\ regions, thereby providing new insights into \textcolor{black}{Hot Molecular Core (HMC) phase} of high-mass star formation and offering implications for stellar feedback within young protoclusters.

\section{The Sample and Observations}
\label{Sample}

\begin{figure*}[!hbt]
    \centering
    \includegraphics[width=18cm]{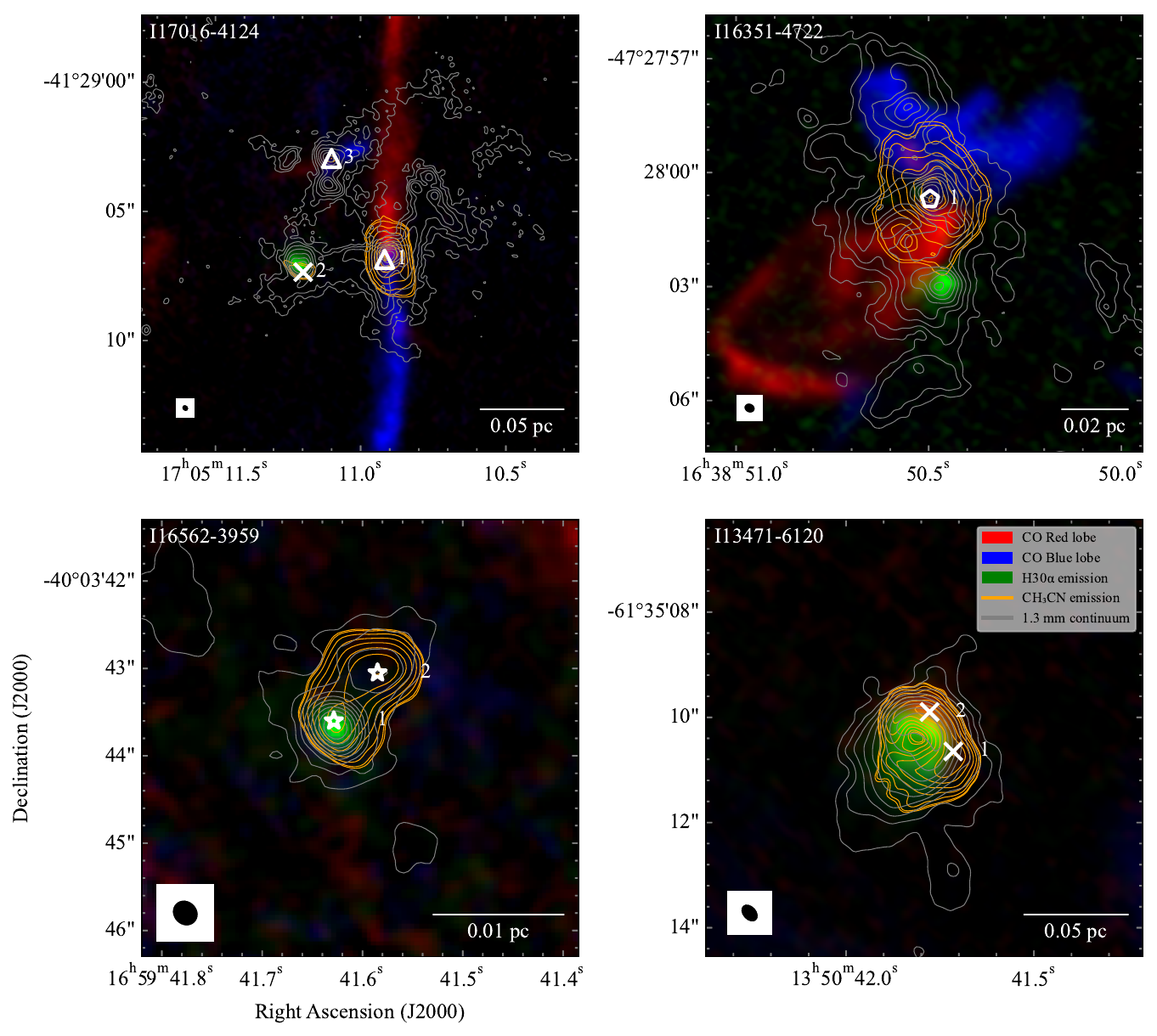}
    \caption{Images of the hydrogen  recombination lines emssion and the outflow for four exemplar sources. The background shows the three-color image composed by the red-shifted \CO\ (red), H30$\alpha$ integrated line emission (green) and the blue-shifted \CO\ (blue). The gray contours and orange contours represent the 1.3 mm continuum and the \ch3cn\,(12$_3$--11$_3$) integrated line emission, respectively. 
    H30$\alpha$ integrated velocity range is [Vlsr-40\,\kms, Vlsr+40\,\kms], where Vlsr is the central velocity (Table \ref{tab:hii} column (9)). For \ch3cn\,(12$_3$--11$_3$), the integrated velocity ranges are different for different fields. 
    The contour levels were plotted from 3\,\rms\ to the peak intensity of the field, with 8 logarithmically spaced contours between these values.
    White markers mark the positions of different type of \textcolor{black}{HMFs}: triangle (jet-like outflow), pentagon (wide-angle outflow), star (no/weak outflow) and cross (shell-like shape). 
    The synthesized beams are shown in the lower left corner, and the scale bar is indicated in the lower right corner of each image.
    }
    \label{fig:example}
\end{figure*}

\begin{figure*}[!hbt]
    \centering
    \includegraphics[width=18cm]{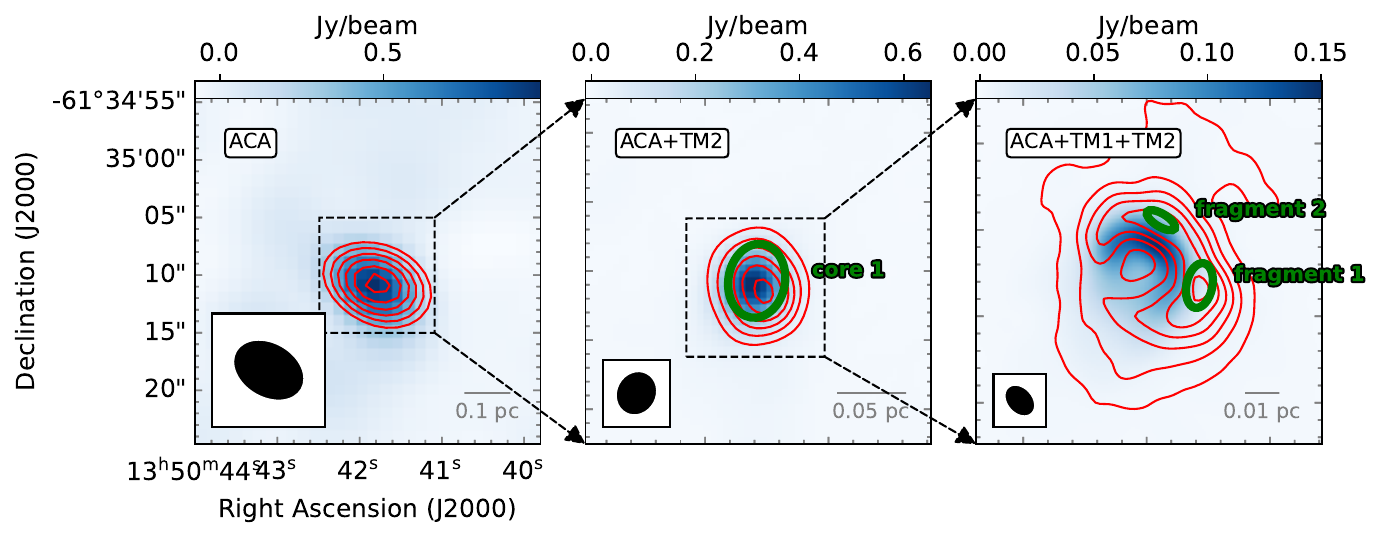}
    \caption{Example of \astrodendro\ results at different scales. The left, middle, and right panels show data from ACA \citep{quarks2}, ACA+TM2 \citep{quarks3}, and ACA+TM1+TM2 \citep{LXC2024}, respectively. The background shows the 1.3 mm continuum. The red contours represent the integrated intensity of \ch3cn\,(12–11), with contour levels ranging from 3 \rms\ to 0.95 $\times$ the peak value, evenly spaced into five levels. The green ellipses outline the structures identified by \astrodendro{}. The synthesized beams (left panel: $\sim$5\arcsec; middle panel: $\sim$ 1\arcsec; right panel: $\sim$ 0.3\arcsec) are shown in the lower left corner, and the scale bar is indicated in the lower right corner of each panel.
    }
    \label{fig:multiscale}
\end{figure*}

\begin{figure*}
    \centering
    \includegraphics[scale=0.63]{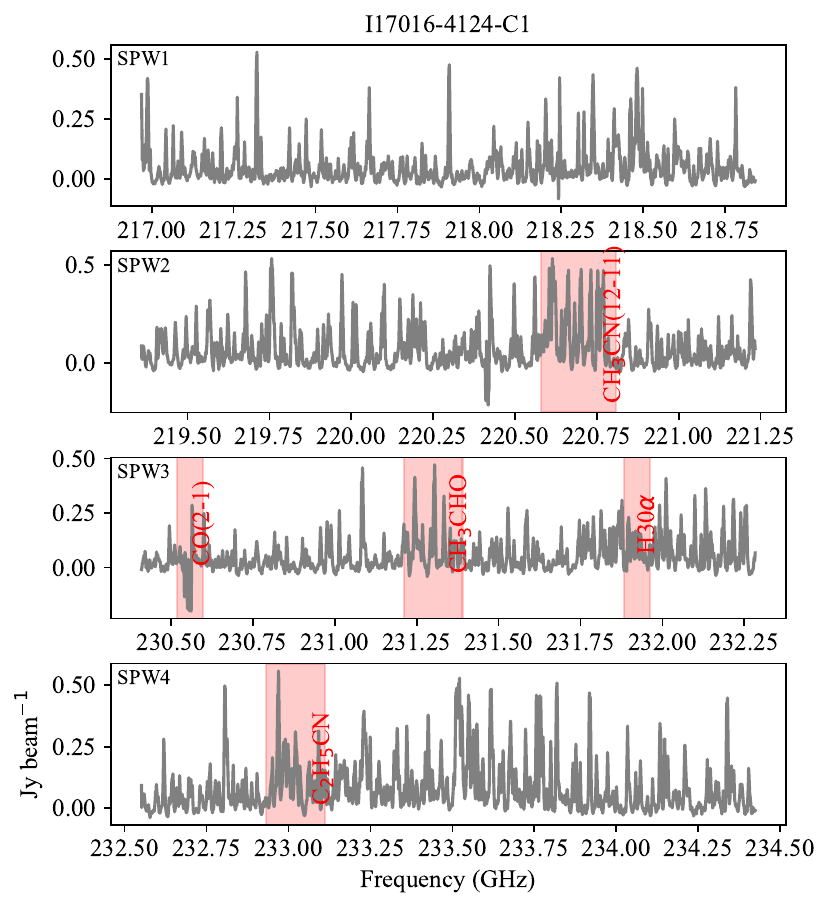} \hfill
    \includegraphics[scale=0.63]{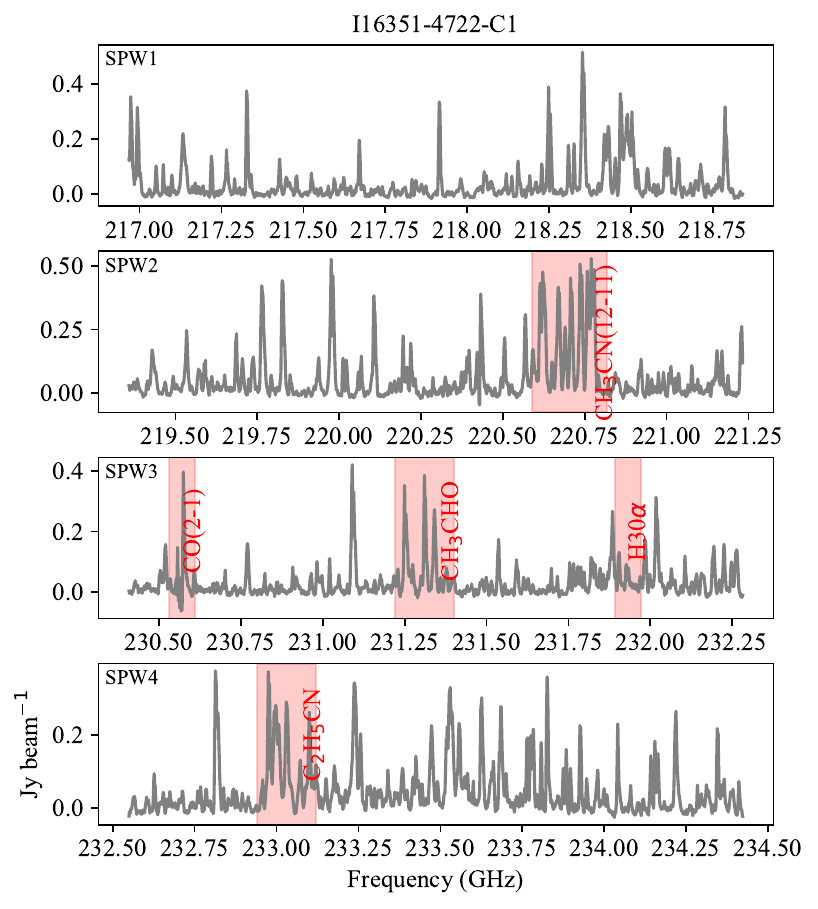} \hfill
    \includegraphics[scale=0.63]{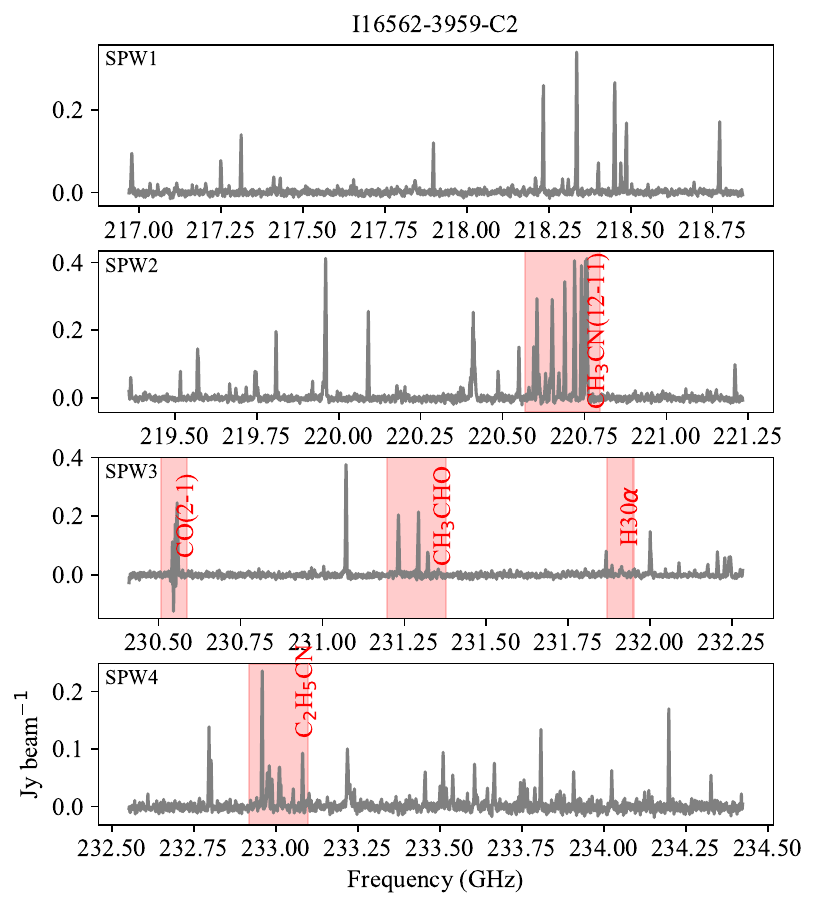} \hfill
    \includegraphics[scale=0.63]{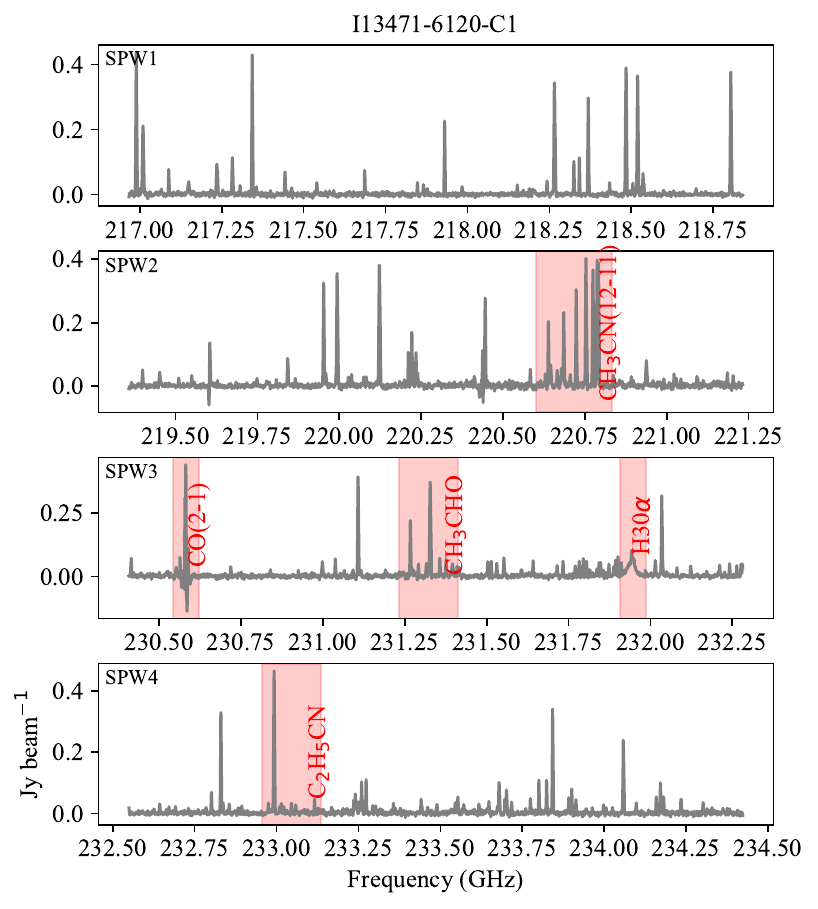}
    \caption{The spectra of all four windows at Band 6 at the \ch3cn\,(12-11) emission map peaks (the positions of the white markers of Figure \ref{fig:example}) for different kinds of \textcolor{black}{HMFs}. Top-left panels: I17016-4124-1; top-right panels: I16351-4722-1; bottom-left panels: I16562-3959-2; bottom-right panels: 13471-6120-1. The positions of \CO{}, \ch3cn\,(12-11), C$_2$H$_5$CN\,(v=0), CH$_3$CHO\,(v=0,1\&2) and H30$\alpha$ are marked with red shadows.
    }
    \label{fig:spectrum}
\end{figure*}


The QUARKS survey observed 139 massive protoclusters in ALMA Band 6. For each source, observations were conducted using the Atacama Compact 7-m Array \textcolor{black}{(ACA)}, ALMA 12-m array in C-2 \textcolor{black}{(TM2)} and C-5 \textcolor{black}{(TM1)} configurations. The combined data from these three configurations provide a resolution of $\sim$\,0.3\arcsec{}, an rms noise (\rms{}) of $\sim$\,3\,mJy\,beam$^{-1}$ for 1.3 mm continuum emission, and a Maximum Recoverable Scale (MRS) of $\sim$27\arcsec. The typical noise level for lines is  $\sim$5 mJy\,beam$^{−1}$ per 0.976 MHz channel ($\sim$1.3 km\,s$^{−1}$). More detailed information on observation and data reduction can be found in \citet{LXC2024} and \citet{quarks2}. In this work, we use the 1.3 mm continuum emission data, as well as line emission data of \ch3cn\,(12--11), H30$\alpha$ (for fields lacking H30$\alpha$ detections, we substitute ATOMS Band 3 H40$\alpha$), and $^{12}$CO (2--1).

From the ATOMS survey, we first selected fields exhibiting H40$\alpha$ emissions from H40$\alpha$ line integrated intensity maps (above 5\,\rms{}). This initial selection ensured that all targets host ionized gas associated with \HII\ regions. Subsequently, by combining QUARKS \ch3cn\,(12--11) channel maps and spectra, we further narrowed down the selection to fields showing \ch3cn\,(12--11) emissions (above 5\,\rms{}) from the initially chosen sample. Through these two-step selection, our sample contains 43 fields, 10 of which lacked H30$\alpha$ detections in the QUARKS Band 6, but were detected in H40$\alpha$ at Band 3.
Table \ref{tab:fields} lists the basic parameters of this sample, including source IDs in our sample (column 1), IRAS names (column 2), coordinates (columns 3-4), systemic velocities ($V_{\mathrm{LSR}}$, column 5), distances from the sun (column 6), Galactocentric distances ($R_\mathrm{GC}$, column 7), effective radius (column 8), dust temperature ($T_\mathrm{dust}$, column 9), bolometric luminosity ($L_{\mathrm{bol}}$, column 10) and clump masses ($M_\mathrm{clump}$, column 11).
Columns (2)-(6) are from \citet{LXC2024}, while columns (7)-(11) are from \citet{LT2020}.
In this sample, different evolutionary stages of high-mass star formation coexist within protoclusters, including hot cores and \HII\ regions. 
The \HII\ regions are identified via hydrogen radio recombination lines (H30$\alpha$ or H40$\alpha$), while the hot cores are traced using \ch3cn\,(12--11) emission.

\textcolor{black}{
For the analysis, we primarily used the H30$\alpha$ where available, given its higher angular resolution in Band~6. In the few cases where H30$\alpha$ data were not available, the H40$\alpha$ was used instead to trace the ionized gas.}

\section{Results}
\label{Results}

Figure \ref{fig:example} presents four examples of sources present in the sample. 
 The \HII\ region candidates, traced by H30$\alpha$, show different evolutionary stages, from compact early phases to expanding cometary structures. 
\12CO\,(2--1) reveal the outflows driven by young stellar objects (YSOs).
The hot core candidates, indicated by \ch3cn\,(12$_3$--11$_3$) emission, exhibit diverse characteristics, with some driving bipolar outflows (e.g., I17016-4124) while others show only weak or non-detectable outflows (e.g., I16562-3959). Zoom-in views of the images for all selected sources are presented in Figure \ref{fig:sample_image}.


\subsection{\textcolor{black}{Identification of Hot Molecular Fragments}}

\textcolor{black}{
Hot molecular cores do not have a clear definition in literature, but are usually referred to gas structures    with size$\lesssim$\,0.1\,pc, number density n$_{\mathrm{H_2}} \gtrsim$ 10$^6~\mathrm{cm}^{-3}$, gas temperature T$>$100 K, and are rich in COMs \citep{Gar1999,May1999,Kur2000,Ces2005,QSL2022,LZY2025}. We use the ACA or ACA+TM2 data ($\sim$1\arcsec{}, or $0.023\ \mathrm{pc}$ at the median distance of our sample, 4.76 kpc) to identify Hot Molecular Cores (HMCs), following the similar procedure in \cite{QSL2022}. These cores correspond to what earlier, lower-resolution studies referred to as hot cores, which likely imply the presence of one or several embedded objects.}

\textcolor{black}{In addition, we use the ACA+TM1+TM2 data ($\sim$0.3\arcsec{}, or $0.006\ \mathrm{pc}$ at 4.76 kpc) to identify the internal substructures within HMC—Hot Molecular Fragments (HMFs). It is important to note that our use of the term “HMF” does not necessarily imply the presence of an embedded object; rather, it refers to hot emission structures of molecular gas regardless of whether they are internally or externally illuminated objects \citep{Gar1999}. Therefore, two terms are used in this paper to describe hot molecular structures: HMCs in lower-resolution data and HMFs in higher-resolution data. It should be noted that HMCs and HMFs refer to the same physical entities, observed at different spatial resolutions. Figure~\ref{fig:multiscale} presents an example at different resolutions. The left and middle panels show the HMC identified at lower resolution, while the right panel reveals that this HMC fragments into two HMFs at higher resolution. In the discussion section, we mainly use HMFs for analysis. In the following, we describe the details on identifying HMCs and HMFs.
}

\textcolor{black}{
The characteristic of HMC is its chemical richness, exhibiting emission from numerous COMs. Many studies utilize different COMs to identify HMC. For example, \citet{Bonfand2024} established a comprehensive census of HMC candidates based on the detection of two CH$_3$OCHO emission lines. In this work, we use \ch3cn\,(12-11) to trace and identify HMF and HMC. \ch3cn\,(12-11) has been demonstrated to be a good tracer of HMC by \citet{Rosero2013}.}
\textcolor{black}{HMFs} were extracted from the \ch3cn\,(12$_3$--11$_3$) integrated intensity maps \textcolor{black}{in the ACA+TM1+TM2 combined data} using the \astrodendro\ package\footnote{\url{http://www.dendrograms.org/}}. 
The \astrodendro\ identifies the changing topology of the surfaces as a function of contour levels and extracts a series of hierarchical structures over a range of spatial scales \citep{Astroden2008}.
The \astrodendro\ was adopted with a minimum threshold of 5\,\rms\ to identify structures, a minimum delta of 3\,\rms\ to distinguish individual \astrodendro\ leaves, and a minimum leaf size equal to the number of pixels within one synthesized beam. Across the 43 sources in our sample, a total of 125 \textcolor{black}{HMFs} were identified and listed in Table \ref{tab:hotcores}.
\textcolor{black}{
In parallel, we also applied \astrodendro\ to the ACA+TM2 continuum to identify the associated larger structures-HMCs. These HMCs more closely resemble the definition of hot cores in the general literature. The \astrodendro\ parameters used for identifying the HMCs were identical to those employed for HMFs. The designations of these HMCs are listed in Column (1) of Table \ref{tab:hotcores}. Figure \ref{fig:multiscale} provides a representative example of the identification results for both HMCs and HMFs.}

\textcolor{black}{Merely Detecting \ch3cn\,(12--11) lines cannot guarantee that the objects are real HMCs or HMFs; it is necessary to determine their gas temperatures.} To estimate the temperatures, we applied the non-local thermodynamic equilibrium (non-LTE) radiative transfer and spectral modeling code \radex{}\footnote{\url{https://spectralradex.readthedocs.io/en/latest/}} \citep{Van2007} to \ch3cn\,(12--11) (K = 0-5) spectra extracted by averaging all pixels within one beam around the identified peak of each \textcolor{black}{HMF}. For some sources, the \ch3cn\,(12–-11) spectra exhibit either optical thickness or blended velocity components, making it difficult to constrain reliably the kinetic temperature through spectral fitting. 
Thus, we excluded cases where the fitting yielded unrealistically high temperatures ($>$900\,K, three times of hot core typical temperature 300\,K, \citealt{Beu2025}).
Among all the fitted sources, the calculated kinetic temperature range is from 95\,K to 798\,K with a median value of 209\,K. 
The kinetic temperature and other parameters of the \textcolor{black}{HMFs} derived from spectral fitting are listed in Table \ref{tab:hotcores}.

\textcolor{black}{Hot cores are not only characterized by their high temperature and density, but also by their chemical richness. We present four spectra of all four windows at Band 6 from representative candidates of the four categories of \textcolor{black}{HMFs} (see Figure \ref{fig:spectrum}). In addition to \ch3cn (12-11) transitions, their spectra exhibit numerous other molecular emission lines. In summary, our identification of HMCs and HMFs is accurate and unequivocal.}

\subsection{\textcolor{black}{Identification of \HII\ regions}}

Similarly, the structures of \HII\ region candidates (hereafter, “\HII\ regions” for brevity) were identified using the \astrodendro\ package applied to the H30$\alpha$ line integrated intensity maps, with the same parameter settings as those used to identify \textcolor{black}{HMFs}. 
For 10 fields that show H30$\alpha$ non-detection, we substituted H40$\alpha$ line integrated intensity maps to identify \HII\ regions (giving priority to H30$\alpha$ because of its higher spatial resolution). 
In total, 64 \HII\ regions were identified and are listed in Table \ref{tab:hii}.
To further classify these \HII\ regions, we extracted the spectra averaged over all pixels within each identified leaf and performed Gaussian fitting to the H30$\alpha$ line. The resulting fitting parameters are provided in Table \ref{tab:hii}.
For the sources without detectable H30$\alpha$ emission, we adopted the sizes and line widths derived from H40$\alpha$ instead. A visual inspection of these H40$\alpha$ structures in Figure \ref{fig:hotcores_outflows} shows that the ionized regions lacking H30$\alpha$ detections are generally extended, suggesting that they may correspond to relatively more evolved UC H\,{\sc ii} regions.
Following the criteria of \citet{LHL2021}, we classified the ionized structures into HC and UC H\,{\sc ii} regions based on their measured sizes and line widths (see Figure \ref{fig:HIIcla} and Table \ref{tab:hii} column (12)).
Specifically, we considered ionized structures with line widths greater than 40 \kms\ and sizes smaller than 0.05 pc as candidate HC\,H\,{\sc ii} regions, while those with line widths between 10–40 \kms and sizes larger than 0.05 pc were classified as candidate UC\,H\,{\sc ii} regions.

\subsection{Classification of \textcolor{black}{HMFs}}
\label{Classification of Hot Cores}
We classify the \textcolor{black}{HMFs} based on their associations with  \HII\ regions and outflows. The \textcolor{black}{HMFs} spatially coinciding with \HII\ regions are proposed to be more evolved than those without any \HII\ regions. Outflow collimation and morphology changes with time \citep{Arce2007}, indicating that outflows can be an alternative evolutionary probe of YSOs. The youngest CO outflows are highly collimated, while more evolved outflows show wider opening angles \citep{Arce2007}.
In this paper, we define spatial separation or coincidence as whether the distance between the peaks is larger than the beam size.

First of all, we clarify how the \CO\ outflow maps were obtained. we checked the \CO\ data channel by channel for each fields. The different velocity ranges were selected to integrate the outflow lobes for different fields in order to avoid contamination and clearly reveal the outflow features. 
In the final outflow maps, we define \textcolor{black}{HMFs} with an outflow lobe signal-to-noise ratio greater than 5\,\rms\ as \textcolor{black}{HMFs} with outflows, while those with signal-to-noise ratio below 5\,\rms\ are referred to as \textcolor{black}{HMFs} without detectable outflows or with very weak outflows.

The identified \textcolor{black}{HMFs} can be broadly classified into four categories.
The first category consists of \textcolor{black}{HMFs} that exhibit prominent \CO\ bipolar jet-like outflows and coincide with 1.3\,mm continuum peak, such as \textcolor{black}{I17016-4124-1} (shown in Figure~\ref{fig:example}). 
The second category consists of \textcolor{black}{HMFs} that show wide-angle outflows and coincide with 1.3\,mm continuum peak, such as I16351-4722-1. Note that the distinction between jet-like and wide-angle outflows is made only by visual inspection. Since there are only nine wide-angle cases, in the subsequent analysis we combine the jet-like and wide-angle types together and refer to them as \textcolor{black}{HMFs} with outflows. 
Here we simply emphasize the existence of such a subtype of \textcolor{black}{HMFs} with wide-angle in our sample.
The third category comprises \textcolor{black}{HMFs} without detectable outflows or with very weak outflows but coinciding with 1.3\,mm continuum peak (such as I16562-3959-2 shown in Figure~\ref{fig:example}). 
The fourth category includes \textcolor{black}{HMFs} located around \HII\ regions, which lack \CO\ outflows and show a spatial separation between the \ch3cn\ emission peak and the 1.3 mm continuum peak, such as in I13471-6120-1/2 (shown in Figure~\ref{fig:example}). Additionally, these \textcolor{black}{HMF} types are categorized in column (14) of Table \ref{tab:hotcores} as: 
(A) jet-like outflow, 35 cores (28\%), 
(B) wide-angle outflow, 9 cores (7\%), 
(C) no/weak outflow, 33 cores (26\%), 
(D) shell-like, 48 cores (38\%, Note that not all \textcolor{black}{HMFs} in this category exhibit a shell-like morphology. Their common characteristic is the lack of spatial coincidence with the 1.3 mm continuum emission, rather than their specific morphology).
The first three types (A, B, C) of \ch3cn\ emission are spatially coincident with the continuum emission, indicating that the hot molecular gas is centered on a strong compact continuum source within the \textcolor{black}{fragment}. These \textcolor{black}{HMFs} are consistent with those commonly described in other literature, in which the heating possibly is internal \textcolor{black}{(candidate internally heated HMFs)}. In contrast, the fourth type (D) of \ch3cn\ emission is not centered on a strong compact continuum source but rather located adjacent to one. These molecular gas are unlikely to be internally heated and are tentatively referred to here as \textcolor{black}{candidate} externally heated \textcolor{black}{HMFs} \citep{ZPT2011}.  \textcolor{black}{They may be residual molecular gas or newly collected gas irradiated by nearby ionized sources}.


\begin{figure*}[!hbt]
    \centering
    \includegraphics[width=18cm]{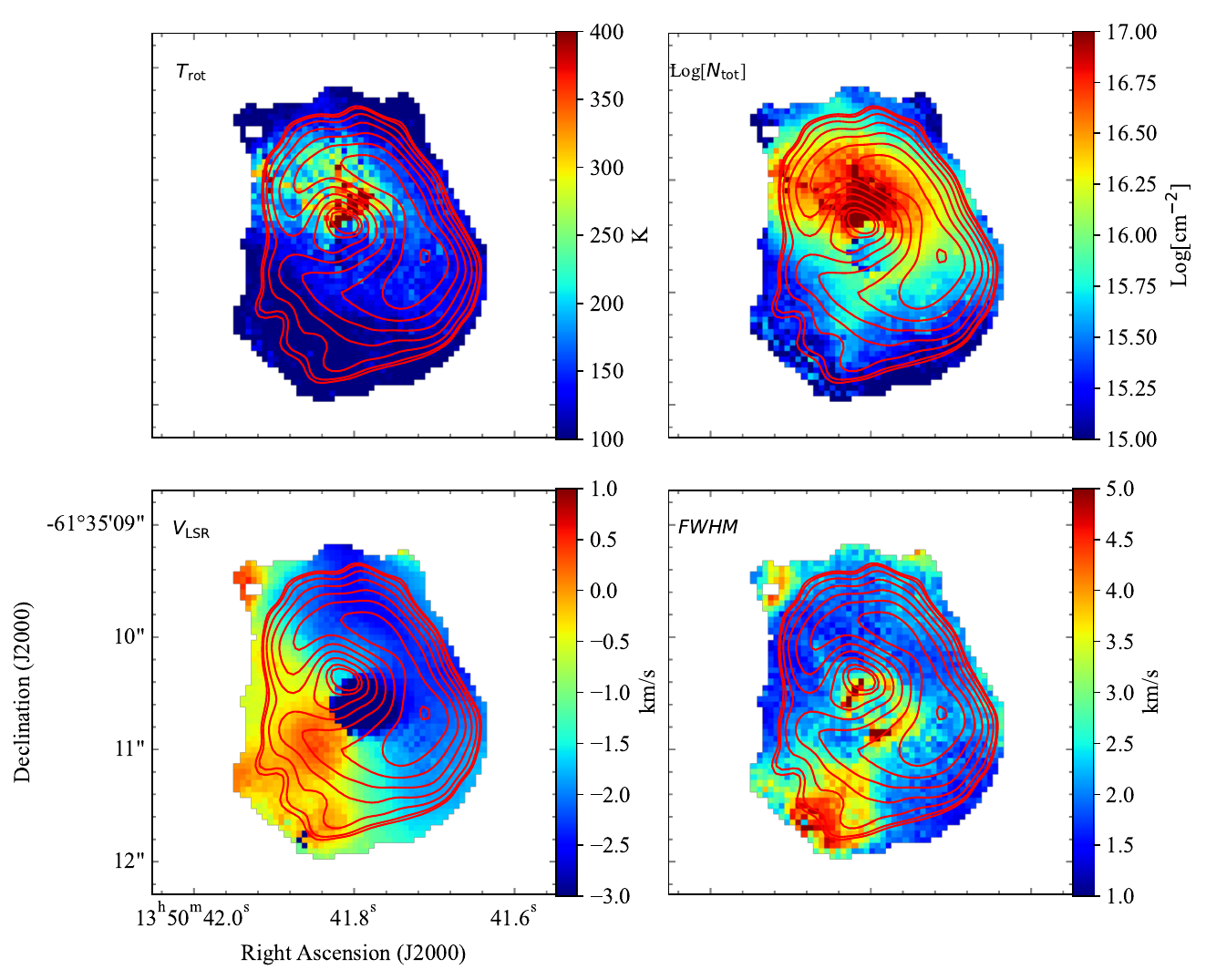}
    \caption{An example result from simultaneously fitting the  multi-transitions of \ch3cn\,(12-11) pixel by pixel on I13471-6120 field using \spectuner{}. Upper panels: rotational temperature (left) and maps of column density  (right). Lower panels: maps of $V_\mathrm{LSR}$ (left, the system velocity has already been deducted) and line width  (right).
    The red contours are the \ch3cn\,(12$_3$--11$_3$) integrated line emission, and the levels are from 3\,\rms\ to the peak intensity of the field, with 8 logarithmically spaced contours between these values.
    }
    \label{fig:mapfit}
\end{figure*}

\begin{figure*}[!bht]
    \centering
    \includegraphics[width=18cm]{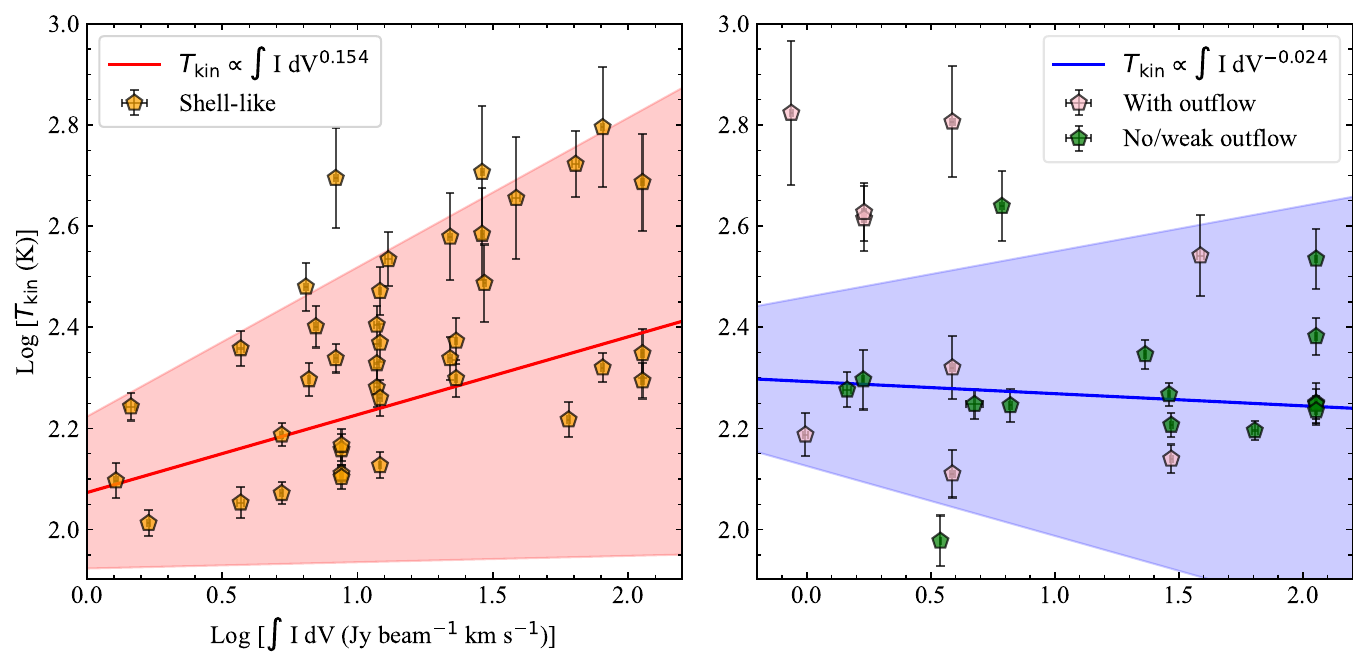}
    \caption{Kinetic temperature of \textcolor{black}{HMFs} versus H30$\alpha$ $\int I \mathrm{d} V$ of the nearest \HII\ regions. Left panel: shell-like \textcolor{black}{HMFs} (orange pentagons). Right panel: \textcolor{black}{candidate} internally heated \textcolor{black}{HMFs} with outflows (pink pentagons) and no/weak outflow (green pentagons). A linear regression is applied to shell-like and \textcolor{black}{candidate} internally heated \textcolor{black}{HMFs}. The corresponding relations $T_\mathrm{kin}\propto\,\int I \mathrm{d} V^{~0.154}$ and $T_\mathrm{kin}\propto\,\int I \mathrm{d} V^{-0.024}$  are shown with red and blue solid lines. The red and blue shadows show 3$\sigma$ uncertainties of  fitting parameters.}
    \label{fig:TvsEM}
\end{figure*}

\begin{figure*}
    \centering
    \includegraphics[scale=0.8]{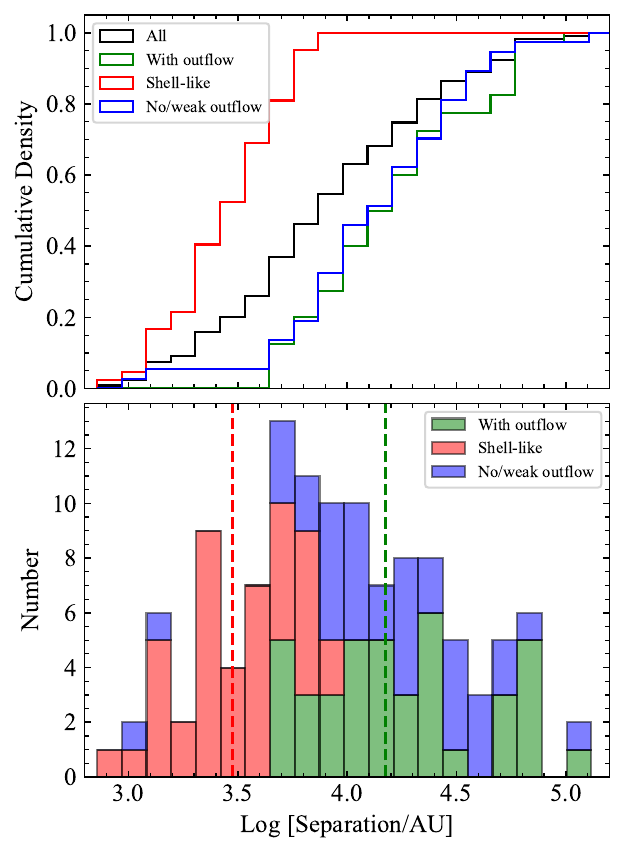} \hfill
    \includegraphics[scale=0.8]{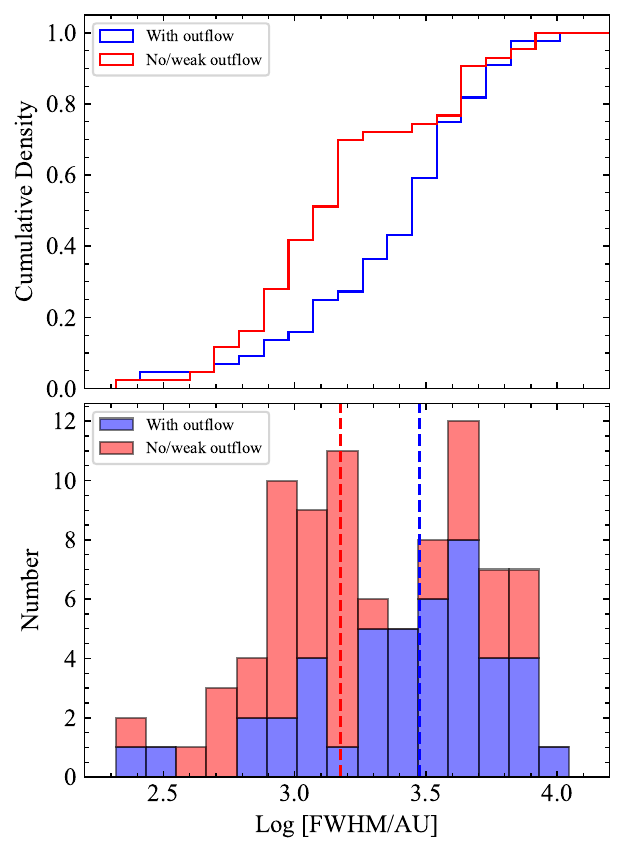}
    \caption{Left column: histogram (bottom) and cumulative density distribution (top) for the separation between different kinds of \textcolor{black}{HMFs} and the nearest \HII\ regions.
    For the shell-like type, the nearest \HII\ region is defined as the \HII\ region enclosed by the \textcolor{black}{nearby} \ch3cn\ emission, whereas for the other types, the nearest \HII\ region is taken to be the closest \HII\ region within the entire image. (Figure \ref{fig:hotcores_outflows}).
    The red and green vertical dashed lines indicate the median value of externally (shell-like) and internally \textcolor{black}{candidate} (with and without outflow) heated \textcolor{black}{HMFs}, respectively.
    The separation is calculated using the distance from \citet{LXC2024} and the intensity peak coordinates of Table \ref{tab:hotcores}, \ref{tab:hii}. Right column: histogram (bottom) and cumulative density distribution (top) for the FWHM ($=\sqrt{\mathrm{Maj\_FWHM}\times \mathrm{Min\_FWHM}}\times \mathrm{distance}$) of different kinds of \textcolor{black}{HMFs}.
    The red and blue vertical dashed lines indicate the corresponding median value.
    }
    \label{fig:hist}
\end{figure*}

\begin{figure*}[!htb]
    \centering
    \includegraphics[width=18cm]{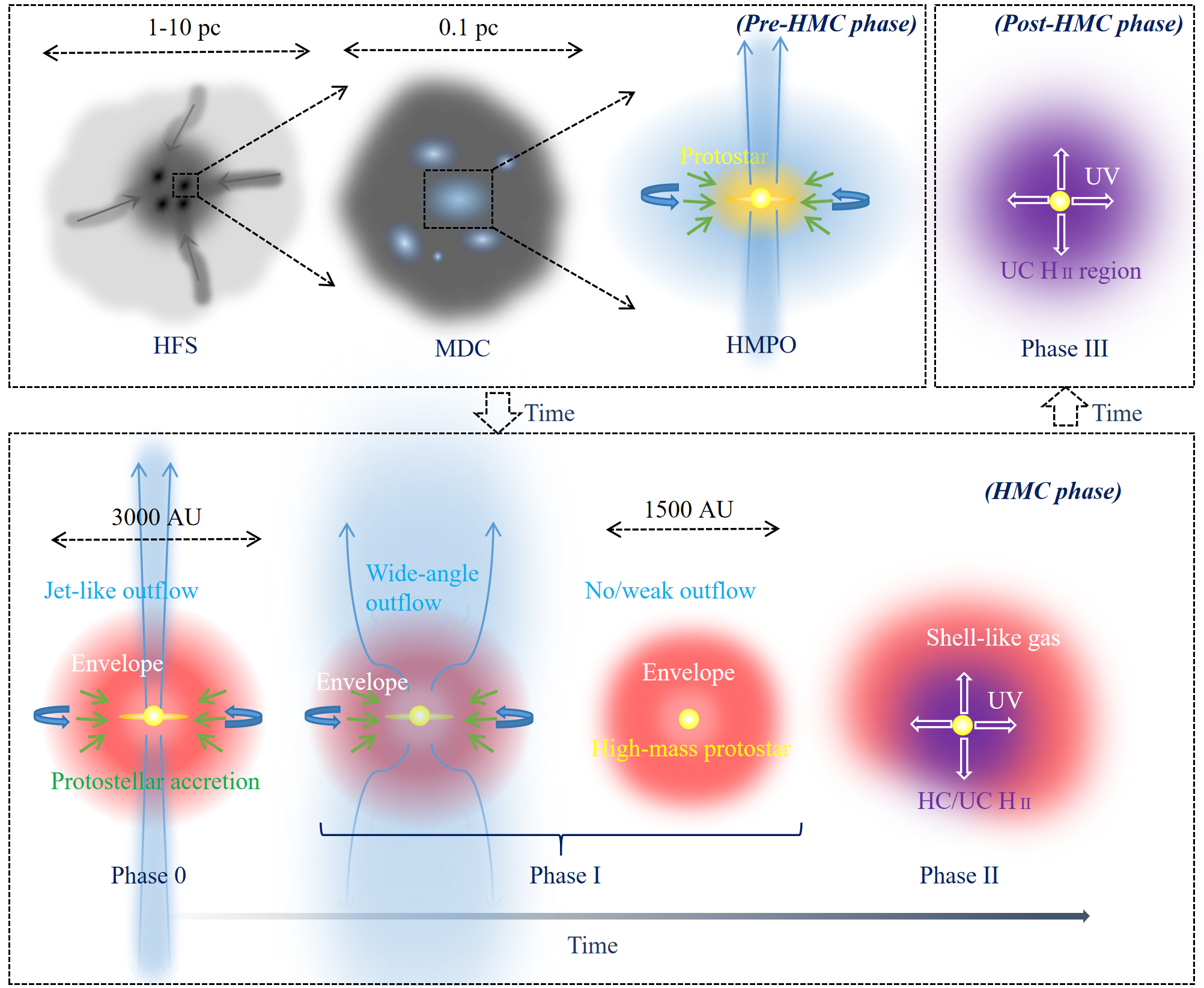}
    \caption{An illustrated overview of high-mass protostellar evolutionary classes. HFS: hub-filament system, the primary birthplace of massive stars. MDC: massive dense core, formed via fragmentation in hub, center undergoes further internal fragmentation to produce multiple prestellar cores. HMPO: high-mass protostellar objects, with jet-like outflow \& without hot \textcolor{black}{molecular} envelope.  Phase 0: jet-like outflow \& hot \textcolor{black}{molecular} envelope (e.g., I17016-4124-1); Phase I: wide angle outflow (e.g., I16351-4722-1) or no outflow (e.g., I16562-3959-2) \& hot \textcolor{black}{molecular} envelope; Phase II: \HII\ \& shell-like \textcolor{black}{HMFs} (e.g., 13471-6120-1/2); Phase III: \HII\ without hot \textcolor{black}{molecular} envelope.
    The gray, blue, red and purple represent cold ($\lesssim$20\,K), warm  (20$\sim$100\,K), hot ($\gtrsim$100\,K) and ionized regions, respectively.
    }
    \label{fig:model}
\end{figure*}


\section{Discussion}
\label{Discussion}

\subsection{Heating Mechanisms of \textcolor{black}{HMFs} \textcolor{black}{adjacent to} \HII{} regions}
\label{Heating Mechanisms of Hot Molecular Gas}


The shell-like \textcolor{black}{HMFs} are not centered on a strong, compact continuum source. If the heating were internal, the hot molecular gas would be expected to be centered on a strong, compact continuum source within the core; however, this is not the case. Instead, the observed COM (\ch3cn{}) emission appears \textcolor{black}{adjacent to} the continuum source. This indicates that it is unlikely to be heated from inside, but rather is more consistent with being heated externally \citep{ZPT2011}.

To further investigate the heating mechanisms of these shell-like \textcolor{black}{HMFs}, we examined the relationship between \textcolor{black}{HMFs} temperatures ($T_{\rm kin}$) and the H30$\alpha$ line integrated intensity ($\int I \mathrm{d} V$) of their nearest \HII\ regions (Figure~\ref{fig:TvsEM}).
Since the line integrated intensity is proportional to the surface H30$\alpha$ line luminosity (i.e., $\int I \mathrm{d}V \propto L$(H30$\alpha$)$/4\pi R^2$, where $R$ is the radius of the \HII\ region), it represents the radiative power per unit area.
So the analysis in this section excludes 10 fields that have non-detectable H30$\alpha$ emission and those \ch3cn\ coincide with H30$\alpha$. 
The shell-like \textcolor{black}{HMFs} exhibit a statistical power-law correlation between $T_{\rm kin}$ and $\int I \mathrm{~d} V$, which is stronger than the non-significant trend observed for internally heated sources (see Figure \ref{fig:TvsEM}). 
We performed a power-law fit between the \textcolor{black}{HMFs} temperatures ($T_{\rm kin}$) and the integrated intensities of the H30$\alpha$ line using the Orthogonal Distance Regression (ODR) package in scipy. In the fitting process, the uncertainties in both variables were taken into account. The best-fit power-law indices are 0.154$\pm$0.047 for the shell-like type and -0.024$\pm$0.038 for the other types.
This result suggests that these shell-like \textcolor{black}{HMFs} are likely heated by radiation from nearby \HII\ regions, whereas the temperature of the \textcolor{black}{candidate} internally heated \textcolor{black}{HMFs} appears to have no connection with the \HII\ regions \textcolor{black}{located near} them.
\textcolor{black}{\citet{ZPT2011} suggested that the “hot core” in Orion KL is primarily shock-heated. For shell-like HMFs in this work, we do not exclude the possibility of shock heating or other types of feedback. The specific feedback mechanisms likely vary from source to source and require detailed studies on individual cases.}

Future investigations are underway to fully resolve the gas kinematics and temperature distributions of these shell-like (externally heated) \textcolor{black}{HMFs} using \spectuner{}\footnote{\url{https://spectuner.readthedocs.io/en/latest/?badge=latest}}, which implements the one-dimensional LTE spectral line model \citep{Qiu2025}.
Here, we present a \spectuner{} \ch3cn\,(12-11) fitting result for the shell-like source I13471-6120 (see Figure \ref{fig:mapfit}). The temperature map shows that the highest temperature region is enclosed by the shell-like \ch3cn\ emission, suggesting that the hot molecular gas traced by the \ch3cn\ shell may indeed be heated by the \textcolor{black}{nearby} ionized region. 
\textcolor{black}{
It should be noted that the term “externally heated” here refers specifically to the HMFs (Figure~\ref{fig:multiscale}, right panel), rather than to the larger-scale HMC associated with them (Figure~\ref{fig:multiscale}, middle panel).}
Furthermore, the velocity map reveals a gradient from southeast to northwest, which is roughly aligned with the direction of the associated cometary ionized gas, indicating that this gradient is likely driven by the expansion of the ionized gas.

Therefore, consistent with the co-evolutionary scenario of HMCs and \HII\ regions proposed in previous studies (e.g., \citealt{Zin2007,Tan2014,Mot2018}), as the high-mass protostar forms, the \HII\ region develops but remains confined to the protostellar vicinity, embedded within the hot molecular envelope (e.g., \textcolor{black}{I16562-3959-1}). As the ionized gas begins to expand, it disrupts and eventually breaks through the hot molecular envelope. The expanding \HII\ region then compresses the disrupted molecular gas into a shell-like morphology, forming externally heated \textcolor{black}{HMFs} (e.g., \textcolor{black}{I13471-6120-1/2}).

\subsection{Hot cores are a long-standing phenomenon in the evolution of massive protostars}

The evolutionary sequence of high-mass star formation has been extensively discussed in the literature from an observational  perspective \citep{Purcell2006,Beu2007,Mot2018,Gie2023}. In this framework, High-Mass Protostellar Objects (HMPOs) form within Massive Dense Cores (MDCs). These HMPOs accrete surrounding material and drive bipolar molecular outflows. As the protostar continues to heat its envelope to temperatures above 100\,K, Hot Molecular Cores (HMCs), or hot cores, develop. 
The massive protostar reaches the main sequence and begins emitting powerful Lyman continuum energetic enough to ionize its surroundings, leading to the formation of  a hyper-compact H\,{\sc ii} (HC\,H\,{\sc ii}) region (e.g., \citealt{YAY2019, YAY2021}). 
However, whether HMCs constitute a specific evolutionary stage—i.e., a distinct, short-lived step between the HMPO and \HII\ phases—remains debated.

Historically, many observational studies—particularly single-dish surveys—have treated HMCs as precursors to H\,{\sc ii} regions in the high-mass star-formation sequence \citep{Hatchell1998,Ces2005,Purcell2009}. 
\textcolor{black}{
These low-resolution observations of HMCs mostly exhibit smooth, elliptical morphologies (such as Figure~\ref{fig:multiscale}, middle panel) and cannot resolve their internal substructures. Therefore, the spatial distribution of HMCs alone cannot be used to determine their specific evolutionary stage. However, high-resolution observations of these HMCs can resolve their internal structures (HMFs), whose spatial distribution (i.e., whether they are coincident with the continuum or H30$\alpha$ emission) can be used to assess whether the central object has begun to disperse its envelope (such as Figure~\ref{fig:multiscale}, right panel). Moreover, the driving source of the outflow can be accurately located, and the outflow opening angle can then be used to infer the evolutionary stage of the embedded object \citep{Arce2007}.
In our sample, these HMFs exhibit diverse characteristics, as illustrated by the examples shown in Figure~\ref{fig:example}.
}
\textcolor{black}{I17016-4124-1} exhibits a jet-like outflow and an extended hot envelope, which may be characteristic of the HMPO phase. Compared to it, \textcolor{black}{I16351-4722-1} drives a wide-angle outflow, suggesting that such sources could be at the end of the HMPO stage.
In \textcolor{black}{I16562-3959-2}, no outflow is detected, which may indicate a more evolved state.
In \textcolor{black}{I16562-3959-1}, ionized gas appears within the inner envelope, which could signal the onset of the \HII\ region.
Finally, in I13471-6120, the \HII\ region may have ionized the inner hot gas, forming shell-like structures such as \textcolor{black}{I13471-6120-1} and \textcolor{black}{I13471-6120-2}, which could be externally heated and analogous to the “hot core” found in Orion KL \citep{ZPT2011}.
Overall, we note that this represents only one possible evolutionary scenario inferred from our observations.

To further investigate the evolutionary sequence of \textcolor{black}{candidate} internally heated \textcolor{black}{HMFs}, we compared the sizes of \textcolor{black}{HMFs} with and without outflows across our sample (see the right column of Figure~\ref{fig:hist}, the sizes are calculated from \ch3cn\,(12$_3$--11$_3$) line integrated intensity maps using \astrodendro{}). The Kolmogorov-Smirnov (K-S) test (p-value = 0.0002 $\ll$ 0.003) reveals a statistically significant difference between \textcolor{black}{HMFs} sizes with and without outflows. In addition, the Anderson–Darling (A-D) test, \textcolor{black}{which evaluates whether a sample comes from a specified distribution by giving more weight to the tails,} was also performed for the \textcolor{black}{HMFs} sizes with and without outflows, yielding a significance level of p-value = 0.005. We find that \textcolor{black}{HMFs} without outflows have slightly smaller sizes (median FWHM$\simeq$1500$\pm$420\,AU) than those with outflows (median FWHM$\simeq$3000$\pm$440\,AU).
To eliminate potential distance effects on the above conclusion, we examined the correlation between the \ch3cn\,(12$_3$--11$_3$) line integrated intensity of all \textcolor{black}{HMFs} and their distances. A Pearson correlation test yields $r = -0.038$ with a $p$-value of 0.67, indicating no statistically significant correlation. This result confirms that distance does not bias our findings.

\textcolor{black}{
Compared to low-resolution observations, high-resolution data resolve the internal structures of HMCs and accurately locate the outflows driven by the embedded objects. We therefore suggest that it is necessary to reconsider the HMC phase within the evolutionary sequence of high-mass protostars.
According to the previously proposed evolutionary scenario \textcolor{black}{(e.g., \citealt{Zin2007,Tan2014,Mot2018})}, we present the possible characteristics of the HMC phase—as illustrated in Figure~\ref{fig:model}. For completeness, we depict the entire evolutionary sequence of high-mass star formation, including the earlier hub–filament system phase (e.g., \citealt{atoms15,Dilda2025,MDZ2025}).}
We divide \textcolor{black}{HMCs} phase into four classes:
Phase 0 – characterized by jet-like outflows and hot core envelopes (e.g., \textcolor{black}{I17016-4124-1});
Phase I – exhibiting wide-angle outflows  (e.g., \textcolor{black}{I16351-4722-1}) or lacking detectable outflows (e.g., \textcolor{black}{16562-3959-1/2}), still with hot \textcolor{black}{molecular} envelopes;
Phase II – associated with \HII\ regions and hot \textcolor{black}{molecular} shells (e.g., \textcolor{black}{I13471-6120-1/2});
Phase III – showing \HII\ regions without any detectable hot \textcolor{black}{molecular envelopes}.
It should be noted that this cartoon illustrates the evolutionary path of most high-mass protostars. In several sources from our sample, an \HII\ region has already formed, yet a bipolar outflow is still being driven. These exceptional cases are not illustrated in the cartoon.
We emphasize that this sequence should be regarded as an interpretative framework informed by both the literature and our data, rather than a definitive evolutionary model established solely by the present analysis.

In summary, \textcolor{black}{hot cores} could be a long-standing phenomenon in the evolution of massive protostars. They can appear as very young protostars with outflows, evolved protostars without outflows, or externally heated shells.

\subsection{Implications for stellar feedback within Young protoclusters}

\citet{LHL2021} identified through the ATOMS survey that 12\% (53/453) of compact dense cores simultaneously host \HII\ regions and COMs emission.
Notably, within these sources, hot cores were found to persist for over half of the lifetime of the \HII\ regions, even after the central objects had begun ionizing their surroundings. However, due to the limited angular resolution of the ATOMS, it remained unclear whether these COM emissions originated from residual shell gas \textcolor{black}{located near} the \HII\ regions or a genuine internally heated hot core at a distinct evolutionary stage \citep{JS2025}.

Since this section focuses on the spatial separations between \ch3cn\ emission and H30$\alpha$/H40$\alpha$ emission, we explicitly exclude the sources in which \ch3cn\ emission is spatially coincident with the H30$\alpha$/H40$\alpha$ emission. The definitions of separation and coincidence are provided in Section \ref{Classification of Hot Cores}.

Our QUARKS sample offers sufficient resolution ($\sim$0.3\arcsec, $\sim$1300\,AU at the median distance 4.76\,kpc of our sample) to resolve COM emission adjacent to the \HII\ regions. We find that these COMs around \HII\ regions consist of two distinct types: externally heated, shell-like \textcolor{black}{HMFs} (associated with adjacent \HII\ regions, forming Phase II), and \textcolor{black}{candidate} internally heated \textcolor{black}{HMFs} (including Phase 0 and Phase I). 
For the latter case, it is important to assess whether their formation has been triggered by the nearest \HII\ region through the “collect-and-collapse” process, in which stellar feedback from massive stars strongly influences the \textcolor{black}{nearby} interstellar medium and regulates subsequent star formation. Specifically, the expansion of an \HII\ region can sweep up the ambient molecular gas into a dense shell, which may eventually become gravitationally unstable and fragment to form new stars \citep{Elmegreen1977,Whitworth1994}.
Alternatively, these \textcolor{black}{HMFs} may simply represent independent star-forming sites that happen to be located nearby, having originated from the fragmentation of a common parental clump but evolving at different rates.
However, for the \textcolor{black}{candidate} internally heated \textcolor{black}{HMFs}, we found no evidence of interplay with the nearby \HII\ regions. The separation between these \textcolor{black}{candidate} internally heated \textcolor{black}{HMFs} and their nearest \HII\ regions is 15000$\pm$3600 AU (median value), which is significantly larger than the 3000$\pm$380 AU (median value) separation observed for the shell-like \textcolor{black}{HMFs} (as shown in Figure~\ref{fig:hist} left column). Furthermore, the A-D (p-value $<$ 0.001) and K-S (p-value $<$ 0.001) tests  indicates that the separations between shell-like \textcolor{black}{HMFs} and nearby \HII\ regions (red line in Figure \ref{fig:hist} left column) and those between \textcolor{black}{candidate} internally heated \textcolor{black}{HMFs} and nearby \HII\ regions (black and green line in Figure \ref{fig:hist} left column) are drawn from different distributions. 
Assuming the observed separation for \textcolor{black}{candidate} internally heated \textcolor{black}{HMFs} is attributed to feedback from \HII\ regions, the dynamical timescale of the \HII\ regions can be estimated at $t_{\mathrm{dyn}} \sim \frac{\mathrm{Separation}}{V_S} \lesssim \frac{15000 \mathrm{~AU}}{10 \mathrm{~km} \mathrm{~s}^{-1}}<1 \times 10^4 \mathrm{~yr}$ \citep{LT2017}. However, this is significantly shorter than \textcolor{black}{the lifetime of High-mass prestellar cores ($\sim 1-7\times10^4 \mathrm{~yr}$, \citealt{Mot2018}) or IR-quiet high-mass protostars ($\sim 2\times10^5 \mathrm{~yr}$, \citealt{Mot2018}), which are precursors before developing HMCs,} suggesting that \HII\ feedback alone can not fully explain the formation and evolution of these \textcolor{black}{candidate} internally heated \textcolor{black}{HMFs}.

We propose that the separations between these \textcolor{black}{candidate} internally heated \textcolor{black}{HMFs} and the \HII\ regions are likely governed by turbulent or thermal fragmentation processes \citep{LT2017}. The separations between majority of the \textcolor{black}{candidate} internally heated \textcolor{black}{HMFs} and the nearest \HII\ regions span 4000-60000 AU, covering the characteristic fragmentation scales observed across low-mass ($\sim$\,20000 AU), intermediate-mass ($\sim$\,6000 AU), and high-mass ($\sim$\,4000 AU) regions as reported in \citet{Beu2025}. 
Therefore, sequential high-mass star formation within young protoclusters could result primarily from the fragmentation of the parental clump, rather than being triggered by feedback mechanisms.

\section{Conclusions}
\label{Conclusions}

We present high-resolution images of protoclusters from the QUARKS survey. These images simultaneously reveal \ch3cn\ emission and H30$\alpha$/H40$\alpha$ emission. 
By combining CO outflow and 1.3~mm continuum, we find that the \ch3cn{}-detected structures \textcolor{black}{(Hot Molecular Fragments, HMFs)} can be grouped into categories associated with jet-like outflows, wide-angle outflows, no/weak outflows, and shell-like morphologies. 
\textcolor{black}{These HMFs represent the substructures of Hot Molecular Cores (HMCs), and such classification provides further insight into the nature of HMCs—a long-standing phenomenon in the evolution of massive protostars.}

These images also allow us to resolve \ch3cn\ emission in the vicinity of the H30$\alpha$/H40$\alpha$ regions. We find that the \ch3cn\ emission near the ionized gas in two distinct configurations: residual molecular gas adjacent to the ionized region, and a separate source at a different evolutionary stage. An analysis of the projected separations shows that the former has a median distance of $\sim$3000\,AU, while the latter has a larger median distance of $\sim$15000\,AU, suggesting that the latter sources may not be physically related to the nearby \textcolor{black}{\HII\ region}.

\section*{Acknowledgements}
We thank the referee for providing constructive comments that substantially improved the quality of the paper. 
This work was main funded by National Key R\&D Program of China under grant Nos. (2022YFA1603100 and 2023YFA1608002) and Tianshan Talent Training Program 2024TSYCTD0013. It was also partially funded by the Regional Collaborative Innovation Project of Xinjiang Uyghur Autonomous Region grant 2022E0105, the NSFC under grants 12173075, 11973076, the CAS Light of West China Program XBZG-ZDSYS-202212, The Tianshan Talent Program of Xinjiang Uygur Autonomous Region under Grant No.2022TSYCLJ0005, the Youth Innovation Promotion Association CAS.
Tie Liu acknowledges the supports by National Natural Science Foundation of China (NSFC) through grants No.12073061 and No.12122307, the PIFI program of Chinese Academy of Sciences through grant No. 2025PG0009, and the Tianchi Talent Program of Xinjiang Uygur Autonomous Region.
S.-L. Qin is supported by NSFC under No.12033005.
This work was performed in part at the Jet Propulsion Laboratory, California Institute of Technology, under contract with the National Aeronautics and Space Administration (80NM0018D0004);
G.G. and LB gratefully acknowledges support by the ANID BASAL project FB210003; 
SRD acknowledges support from the Fondecyt Postdoctoral fellowship (project code 3220162) and ANID BASAL project FB210003; 
PS was partially supported by a Grant-in-Aid for Scientific Research (KAKENHI Number JP23H01221) of JSPS;
\facilities{ALMA}
\software{Astropy \citep{Astropy2022} 
          CASA \citep{CASA2022},
          CARTA \citep{Carta2021},
          Astrodendro \citep{Astroden2008},
          RADEX \citep{Van2007},
          Spectuner \citep{Qiu2025}}

\bibliography{template}{}
\bibliographystyle{aasjournal}


\appendix

\renewcommand\thefigure{\Alph{section}\arabic{figure}}
\renewcommand\thetable{\Alph{section}\arabic{table}}

\section{The sample selected from the QuARKS survey}

In this section, we present the basic parameters of 43 fields in Table \ref{tab:fields}, including source IDs in our sample (column 1), IRAS names (column 2), coordinates (columns 3-4), systemic velocities ($V_{\mathrm{LSR}}$, column 5), distances from the sun (column 6), Galactocentric distances ($R_\mathrm{GC}$, column 7), effective radius (column 8), dust temperature ($T_\mathrm{dust}$, column 9), bolometric luminosity ($L_{\mathrm{bol}}$, column 10) and clump masses ($M_\mathrm{clump}$, column 11).

\startlongtable
\tabletypesize{\scriptsize}

\begin{deluxetable*}{rlrrrrlllll}
\renewcommand{\arraystretch}{0.8}
\tablecaption{The basic information of 43 fields in our sample \label{tab:fields}}

\tablehead{
\colhead{ID} & \colhead{Target} & \colhead{Glon} & \colhead{Glat} & \colhead{$V_{\mathrm{LSR}}$} & 
\colhead{Distance} & \colhead{$R_\mathrm{GC}$} & \colhead{Radius} & \colhead{$T_\mathrm{dust}$} & 
\colhead{log($L_{\mathrm{bol}}$)} & \colhead{log($M_\mathrm{clump}$)} \\[-7pt]
\colhead{-} & \colhead{-} & \colhead{(\degree)} & \colhead{(\degree)} & \colhead{(\kms)} & 
\colhead{(kpc)} & \colhead{(kpc)} & \colhead{(pc)} & \colhead{(K)} & 
\colhead{($\mathrm{L}_{\odot}$)} & \colhead{($\mathbf{M}_{\odot}$)} 
}
\decimalcolnumbers
\startdata
1 & I12320-6122 & 300.96916 & 1.14564 & -43.2 & 4.17 & 7.2 & 1.00 & 44.6 & 5.6 & 3.0 \\
2 & I12326-6245 & 301.13583 & -0.22582 & -39.5 & 4.21 & 7.2 & 0.83 & 34.2 & 5.4 & 3.5 \\
3 & I12572-6316\_1 & 303.93008 & -0.68782 & 30.4 & 11.63 & 9.8 & 1.63 & 21.5 & 4.6 & 3.9 \\
4 & I13471-6120 & 309.92084 & 0.47748 & -57.8 & 5.17 & 6.4 & 1.01 & 35.1 & 5.3 & 3.4 \\
5 & I15254-5621 & 323.45906 & -0.07914 & -68.6 & 4.42 & 5.7 & 0.89 & 33.5 & 5.1 & 3.1 \\
6 & I15290-5546 & 324.20074 & 0.12008 & -88.3 & 7.94 & 4.9 & 1.80 & 33.5 & 5.7 & 3.8 \\
7 & I15411-5352 & 326.72494 & 0.61576 & -41.8 & 2.41 & 6.9 & 0.57 & 30.5 & 4.5 & 2.7 \\
8 & I15439-5449 & 326.47310 & -0.37656 & -54.8 & 3.40 & 5.9 & 0.69 & 26.8 & 4.4 & 3.0 \\
9 & I15502-5302 & 328.30750 & 0.43085 & -92.4 & 5.49 & 4.6 & 1.66 & 35.7 & 5.8 & 3.7 \\
10 & I15520-5234 & 328.80764 & 0.63243 & -41.8 & 2.56 & 6.2 & 0.67 & 32.2 & 5.1 & 3.2 \\
11 & I15567-5236 & 329.33743 & 0.14749 & -107.5 & 5.22 & 4.4 & 1.31 & 35.4 & 5.7 & 3.5 \\
12 & I16037-5223 & 330.29433 & -0.39406 & -81.3 & 9.04 & 4.9 & 2.15 & 31.4 & 5.6 & 3.8 \\
13 & I16060-5146 & 330.95419 & -0.18248 & -92.1 & 5.39 & 4.5 & 1.24 & 32.2 & 5.8 & 3.9 \\
14 & I16065-5158 & 330.87788 & -0.36624 & -62.5 & 4.33 & 5.2 & 1.41 & 30.8 & 5.4 & 3.7 \\
15 & I16071-5142 & 331.13062 & -0.24240 & -86.5 & 5.29 & 4.5 & 1.21 & 23.9 & 4.8 & 3.7 \\
16 & I16164-5046 & 332.82560 & -0.54900 & -56.7 & 4.02 & 5.4 & 1.37 & 31.4 & 5.5 & 3.7 \\
17 & I16172-5028\_1 & 333.13502 & -0.43180 & -53.3 & 3.78 & 5.4 & 1.51 & 32.0 & 5.8 & 4.0 \\
18 & I16313-4729 & 336.86551 & 0.00213 & -74.1 & 4.99 & 4.4 & 2.06 & 31.0 & 6.7 & 4.7 \\
19 & I16348-4654 & 337.70403 & -0.05342 & -47.8 & 11.33 & 5.4 & 2.40 & 23.6 & 5.4 & 4.4 \\
20 & I16351-4722 & 337.40474 & -0.40206 & -40.8 & 2.33 & 5.7 & 0.69 & 30.4 & 4.9 & 3.2 \\
21 & I16445-4459 & 340.24917 & -0.04582 & -122.2 & 7.72 & 2.8 & 2.54 & 24.6 & 5.0 & 3.9 \\
22 & I16458-4512 & 340.24749 & -0.37413 & -50.9 & 4.19 & 5.1 & 1.42 & 21.4 & 4.5 & 3.6 \\
23 & I16562-3959 & 345.49528 & 1.47062 & -11.3 & 1.37 & 6.1 & 0.72 & 42.3 & 5.7 & 3.2 \\
24 & I17016-4124 & 345.00285 & -0.22408 & -26.8 & 3.16 & 7.0 & 0.75 & 32.0 & 5.3 & 3.8 \\
25 & I17143-3700 & 350.01583 & 0.43248 & -31.7 & 11.71 & 4.7 & 2.95 & 31.0 & 5.6 & 3.8 \\
26 & I17160-3707 & 350.10365 & 0.08132 & -69.4 & 10.28 & 2.7 & 1.69 & 28.5 & 6.0 & 4.1 \\
27 & I17175-3544 & 351.41758 & 0.64492 & -8.7 & 1.32 & 7.0 & 0.35 & 30.6 & 4.8 & 3.1 \\
28 & I17204-3636 & 351.04114 & -0.33570 & -18.1 & 3.27 & 5.1 & 0.60 & 25.8 & 4.2 & 2.9 \\
29 & I17220-3609 & 351.58185 & -0.35181 & -97.2 & 8.05 & 1.3 & 2.41 & 25.4 & 5.7 & 4.3 \\
30 & I17271-3439\_1 & 353.40990 & -0.36021 & -17.2 & 3.66 & 5.3 & 1.34 & 35.0 & 5.6 & 4.0 \\
31 & I17545-2357 & 5.63735 & 0.23748 & 8.8 & 3.00 & 5.4 & 0.87 & 23.7 & 4.1 & 3.1 \\
32 & I17599-2148 & 8.14078 & 0.22409 & 18.6 & 3.44 & 5.4 & 1.15 & 32.0 & 5.2 & 3.4 \\
33 & I18032-2032 & 9.61969 & 0.19657 & 4.4 & 4.80 & 3.4 & 1.27 & 32.1 & 5.4 & 3.5 \\
34 & I18056-1952 & 10.47249 & 0.02752 & 66.4 & 8.44 & 1.6 & 2.32 & 25.1 & 5.7 & 4.4 \\
35 & I18110-1854 & 11.93713 & -0.61570 & 38.5 & 3.17 & 5.1 & 0.87 & 28.9 & 4.8 & 3.2 \\
36 & I18116-1646 & 13.87386 & 0.28067 & 48.8 & 3.86 & 4.6 & 0.99 & 33.8 & 5.1 & 3.1 \\
37 & I18434-0242 & 29.95723 & -0.01729 & 97.5 & 4.76 & 4.7 & 1.48 & 35.5 & 5.7 & 3.6 \\
38 & I18469-0132 & 31.39533 & -0.25773 & 86.6 & 4.91 & 4.7 & 0.68 & 32.2 & 4.8 & 3.0 \\
39 & I18479-0005 & 32.79860 & 0.18954 & 14.6 & 12.87 & 7.5 & 2.45 & 34.2 & 6.1 & 4.2 \\
40 & I18507+0110 & 34.25691 & 0.15530 & 58.2 & 3.23 & 7.1 & 0.44 & 29.2 & 4.8 & 3.2 \\
41 & I19078+0901 & 43.16581 & 0.01086 & 6.2 & 11.49 & 7.6 & 4.26 & 33.3 & 6.9 & 5.0 \\
42 & I19095+0930 & 43.79414 & -0.12749 & 43.8 & 9.07 & 5.8 & 0.64 & 34.9 & 5.1 & 3.1 \\
43 & I19097+0847 & 43.17804 & -0.51881 & 58.2 & 7.83 & 6.2 & 2.01 & 23.3 & 5.0 & 3.8 \\
\enddata
\end{deluxetable*}

\section{Hot core catalog}
\restartappendixnumbering

 All \textcolor{black}{HMFs} identified using \astrodendro\ are listed in Table \ref{tab:hotcores}. \textcolor{black}{Column (1)-(9) are their corresponding larger-scale HMCs ID, HMFs ID,}  \ch3cn\,(12$_3$--11$_3$) line integrated  intensity peak coordinates RA, DEC, major FWHM and minor FWHM in arcsec, major FWHM and minor FWHM in AU, position angle from measured from north to east.  Columns (10)-(13) list the line-of-sight velocities, line widths, kinetic temperatures and their uncertainties, as well as the \ch3cn\ column
densities and their uncertainties. Rows with “-” indicate sources for which the \ch3cn\ fitting yielded $T_{\rm kin}>900$ K. Since we regard such values as unreliable, these fits were discarded and not considered in the analysis presented in Section \ref{Heating Mechanisms of Hot Molecular Gas}. This cutoff of 900 K is adopted because hot molecular cores in the high-mass star formation phase typically have temperatures of about 300 K \citep{Beu2025}. We therefore adopted three times this characteristic value as a tolerance threshold.

\startlongtable
\begin{deluxetable*}{lrlrllrrrrrrrl}
\tabletypesize{\scriptsize}
\renewcommand{\arraystretch}{0.8}
\tablecaption{Physical parameters of the \textcolor{black}{HMFs} \label{tab:hotcores}}
\tablehead{
  \colhead{\textcolor{black}{Core}} & \colhead{\textcolor{black}{Fragment}} & \colhead{RA} & \colhead{DEC} & \colhead{Maj\_FWHM} &\colhead{Min\_FWHM} & 
  \colhead{Maj\_size} &\colhead{Min\_size} &
  \colhead{PA} & \colhead{Vlsr} & \colhead{Linewidth} &
  \colhead{$T_{\rm kin}$} & \colhead{ERR$_{T}$} & \colhead{Flag$^a$} \\[-7pt]
  \colhead{-}      & \colhead{-}  & \colhead{(\degree)} & \colhead{(\degree)} &
  \colhead{(\arcsec)} & \colhead{(\arcsec)} &
  \colhead{(AU)} & \colhead{(AU)} & 
  \colhead{(\degree)} & \colhead{(\kms)} &
  \colhead{(\kms)} & \colhead{(K)} & \colhead{(K)} & \colhead{-}
}
\decimalcolnumbers
\startdata
I12320-6122-C1 & 1 & 188.72195 & -61.66142 & 0.55 & 0.31 & 2302 & 1301 & -152 & -43.55 & 5.35 & 176 & 13 & C \\
I12320-6122-C1 & 2 & 188.72171 & -61.66103 & 0.89 & 0.46 & 3703 & 1902 & 109 & -36.96 & 3.55 & 198 & 15 & D \\
I12326-6245-C1 & 1 & 188.89608 & -63.04233 & 0.48 & 0.24 & 2021 & 1010 & 174 & -39.35 & 3.41 & 223 & 25 & D \\
I12326-6245-C1 & 2 & 188.89688 & -63.04206 & 0.43 & 0.29 & 1819 & 1212 & 82 & -39.01 & 3.28 & 178 & 16 & C \\
I12326-6245-C1 & 3 & 188.89633 & -63.04206 & 0.82 & 0.36 & 3435 & 1516 & -151 & -38.94 & 5.67 & 486 & 107 & D \\
I12326-6245-C1 & 4 & 188.89682 & -63.04181 & 0.53 & 0.24 & 2223 & 1010 & 60 & -37.67 & 4.44 & 197 & 16 & D \\
I12326-6245-C1 & 5 & 188.89587 & -63.04179 & 0.22 & 0.17 & 909 & 707 & 98 & -44.24 & 5.51 & 343 & 47 & C \\
I12326-6245-C1 & 6 & 188.89749 & -63.04174 & 0.34 & 0.26 & 1415 & 1111 & 140 & -38.22 & 5.68 & 241 & 20 & C \\
I12326-6245-C1 & 7 & 188.89614 & -63.04162 & 0.19 & 0.12 & 808 & 505 & 165 & -41.15 & 3.92 & 177 & 12 & C \\
I12326-6245-C1 & 8 & 188.89559 & -63.04117 & 0.31 & 0.26 & 1314 & 1111 & 131 & -40.42 & 4.75 & 172 & 11 & C \\
I12572-6316\_1-C1 & 1 & 195.10084 & -63.54217 & 0.22 & 0.19 & 2512 & 2233 & 102 & 31.84 & 6.12 & 154 & 15 & A \\
I13471-6120-C1 & 1 & 207.67381 & -61.58629 & 0.50 & 0.29 & 2606 & 1489 & 78 & -60.20 & 3.71 & 218 & 14 & D \\
I13471-6120-C1 & 2 & 207.67407 & -61.58608 & 0.36 & 0.14 & 1861 & 744 & 149 & -60.14 & 3.12 & 495 & 113 & D \\
I15254-5621-C1 & 1 & 232.33054 & -56.52287 & 0.79 & 0.26 & 3501 & 1167 & 84 & -66.77 & 3.94 & 379 & 75 & D \\
I15254-5621-C1 & 2 & 232.33046 & -56.52251 & 0.53 & 0.31 & 2334 & 1379 & 96 & -69.38 & 2.59 & 218 & 21 & D \\
I15290-5546-C1 & 1 & 233.22023 & -55.93535 & 0.91 & 0.74 & 7241 & 5907 & 56 & -88.44 & 7.99 & 412 & 61 & A \\
I15290-5546-C1 & 2 & 233.21956 & -55.93518 & 0.38 & 0.36 & 3049 & 2858 & 73 & -90.66 & 7.88 & 424 & 56 & A \\
I15411-5352-C1 & 1 & 236.24839 & -54.03960 & 0.53 & 0.48 & 1272 & 1157 & 145 & -41.68 & 10.80 & 666 & 219 & A \\
I15439-5449-C1 & 1 & 236.95741 & -54.97669 & 0.86 & 0.50 & 2938 & 1714 & 121 & -53.47 & 9.66 & 209 & 31 & A \\
I15502-5302-C1 & 1 & 238.52730 & -53.19496 & 0.17 & 0.17 & 922 & 922 & 102 & -88.25 & 2.60 & 144 & 10 & D \\
I15502-5302-C1 & 2 & 238.52753 & -53.19479 & 0.77 & 0.41 & 4216 & 2240 & 118 & -92.04 & 5.31 & 129 & 4 & D \\
I15502-5302-C1 & 3 & 238.52656 & -53.19485 & 0.34 & 0.12 & 1845 & 659 & 172 & -92.04 & 3.80 & 127 & 7 & D \\
I15502-5302-C1 & 4 & 238.52668 & -53.19475 & 0.19 & 0.14 & 1054 & 791 & 63 & -92.21 & 4.57 & 147 & 11 & D \\
I15520-5234-C1 & 1 & 238.95234 & -52.71864 & 0.41 & 0.29 & 1044 & 737 & 119 & -41.86 & 4.20 & 296 & 32 & D \\
I15520-5234-C1 & 2 & 238.95197 & -52.71858 & 0.58 & 0.46 & 1475 & 1167 & 83 & -43.53 & 4.32 & 234 & 20 & D \\
I15520-5234-C2 & 3 & 238.95073 & -52.71835 & 0.62 & 0.34 & 1597 & 860 & 140 & -42.84 & 2.95 & 113 & 8 & D \\
I15520-5234-C1 & 4 & 238.95266 & -52.71837 & 0.53 & 0.29 & 1352 & 737 & 155 & -40.65 & 4.44 & 166 & 8 & C* \\
I15520-5234-C1 & 5 & 238.95344 & -52.71824 & 0.53 & 0.29 & 1352 & 737 & 97 & -41.98 & 2.84 & 118 & 6 & D \\
I15520-5234-C2 & 6 & 238.95185 & -52.71824 & 0.22 & 0.14 & 553 & 369 & 76 & -44.18 & 1.91 & 228 & 18 & D \\
I15520-5234-C1 & 7 & 238.95291 & -52.71818 & 0.29 & 0.24 & 737 & 614 & -177 & -42.54 & 5.26 & 154 & 8 & D \\
I15520-5234-C1 & 8 & 238.95224 & -52.71803 & 0.22 & 0.17 & 553 & 430 & 48 & -43.50 & 2.39 & 134 & 8 & D \\
I15567-5236-C1 & 1 & 240.13783 & -52.74604 & 0.31 & 0.29 & 1629 & 1503 & 92 & -113.31 & 4.71 & 177 & 20 & C* \\
I15567-5236-C1 & 2 & 240.13804 & -52.74444 & 0.24 & 0.19 & 1253 & 1002 & -138 & -107.11 & 2.41 & 95 & 11 & C \\
I16037-5223-C1 & 1 & 241.90930 & -52.51721 & 0.50 & 0.48 & 4556 & 4339 & -141 & - & - & - & - & C \\
I16037-5223-C1 & 2 & 241.90884 & -52.51721 & 0.46 & 0.31 & 4122 & 2820 & 127 & -78.62 & 13.46 & 209 & 30 & A \\
I16037-5223-C1 & 3 & 241.90836 & -52.51704 & 0.65 & 0.34 & 5858 & 3037 & -156 & -79.00 & 9.02 & 640 & 162 & A \\
I16037-5223-C1 & 4 & 241.90918 & -52.51681 & 0.31 & 0.26 & 2820 & 2387 & 94 & -80.56 & 7.68 & 129 & 14 & A \\
I16060-5146-C2 & 1 & 242.46818 & -51.91606 & 0.41 & 0.22 & 2199 & 1164 & 109 & -88.34 & 2.86 & 129 & 5 & C* \\
I16060-5146-C1 & 2 & 242.46922 & -51.91586 & 0.43 & 0.31 & 2328 & 1682 & -148 & -89.00 & 4.00 & 185 & 10 & C \\
I16060-5146-C1 & 3 & 242.46852 & -51.91557 & 1.10 & 0.58 & 5951 & 3105 & 60 & - & - & - & - & A \\
I16060-5146-C1 & 4 & 242.46962 & -51.91550 & 0.22 & 0.17 & 1164 & 906 & 98 & -95.97 & 5.54 & 384 & 80 & D \\
I16060-5146-C1 & 5 & 242.46915 & -51.91525 & 0.46 & 0.24 & 2458 & 1294 & 157 & -94.23 & 7.62 & 509 & 154 & D \\
I16060-5146-C1 & 6 & 242.46834 & -51.91519 & 0.29 & 0.12 & 1552 & 647 & 73 & -86.01 & 4.30 & 342 & 42 & D \\
I16060-5146-C1 & 7 & 242.46971 & -51.91517 & 0.26 & 0.10 & 1423 & 517 & 84 & - & - & - & - & D \\
I16060-5146-C1 & 8 & 242.46962 & -51.91490 & 0.55 & 0.26 & 2975 & 1423 & 74 & -95.62 & 5.87 & 624 & 170 & D \\
I16060-5146-C2 & 9 & 242.46845 & -51.91482 & 0.96 & 0.70 & 5174 & 3751 & 165 & - & - & - & - & D \\
I16060-5146-C1 & 10 & 242.46926 & -51.91464 & 0.38 & 0.19 & 2070 & 1035 & -154 & -93.95 & 6.40 & 209 & 14 & D \\
I16065-5158-C1 & 1 & 242.58309 & -52.10344 & 0.41 & 0.24 & 1767 & 1039 & 179 & - & - & - & - & C \\
I16065-5158-C1 & 2 & 242.58257 & -52.10335 & 0.72 & 0.19 & 3118 & 831 & -139 & - & - & - & - & C \\
I16065-5158-C2 & 3 & 242.58296 & -52.10211 & 2.02 & 1.68 & 8729 & 7274 & 123 & - & - & - & - & A \\
I16071-5142-C1 & 1 & 242.74909 & -51.84012 & 0.38 & 0.19 & 2031 & 1016 & 102 & -83.68 & 6.48 & 243 & 20 & C \\
I16071-5142-C1 & 2 & 242.74893 & -51.83967 & 1.08 & 0.74 & 5713 & 3936 & 83 & - & - & - & - & A \\
I16164-5046-C1 & 1 & 245.04601 & -50.88833 & 0.65 & 0.48 & 2604 & 1929 & 61 & - & - & - & - & A \\
I16164-5046-C1 & 2 & 245.04605 & -50.88765 & 1.54 & 1.15 & 6173 & 4630 & 115 & - & - & - & - & A \\
I16164-5046-C1 & 3 & 245.04528 & -50.88729 & 0.55 & 0.36 & 2218 & 1447 & -157 & -57.64 & 6.37 & 252 & 24 & D \\
I16172-5028\_1-C1 & 1 & 245.26256 & -50.58708 & 0.72 & 0.43 & 2722 & 1633 & -168 & -53.23 & 5.00 & 165 & 13 & D \\
I16172-5028\_1-C1 & 2 & 245.26206 & -50.58693 & 0.77 & 0.36 & 2903 & 1361 & 90 & - & - & - & - & D \\
I16172-5028\_1-C1 & 3 & 245.26308 & -50.58686 & 0.48 & 0.14 & 1814 & 544 & 126 & -53.71 & 10.00 & 191 & 17 & D \\
I16172-5028\_1-C1 & 4 & 245.26282 & -50.58681 & 0.31 & 0.12 & 1179 & 454 & 46 & -53.71 & 6.78 & 213 & 19 & D \\
I16172-5028\_1-C1 & 5 & 245.26247 & -50.58679 & 0.36 & 0.26 & 1361 & 998 & 92 & -52.55 & 5.61 & 254 & 22 & D \\
I16313-4729-C1 & 1 & 248.72672 & -47.59375 & 0.65 & 0.46 & 3234 & 2275 & 153 & - & - & - & - & B* \\
I16348-4654-C2 & 1 & 249.62130 & -47.01219 & 0.36 & 0.29 & 4079 & 3263 & 145 & - & - & - & - & C \\
I16348-4654-C1 & 2 & 249.62350 & -47.01006 & 1.03 & 0.94 & 11693 & 10605 & 50 & - & - & - & - & B* \\
I16351-4722-C1 & 1 & 249.71040 & -47.46686 & 1.68 & 1.32 & 3914 & 3076 & 124 & - & - & - & - & B \\
I16445-4459-C1 & 1 & 252.02145 & -45.08560 & 0.67 & 0.48 & 5188 & 3706 & 45 & - & - & - & - & A \\
I16458-4512-C1 & 1 & 252.37523 & -45.29565 & 0.55 & 0.36 & 2313 & 1508 & 67 & -51.44 & 4.00 & 240 & 21 & A* \\
I16562-3959-C1 & 1 & 254.92345 & -40.06211 & 0.19 & 0.12 & 263 & 164 & 78 & -16.95 & 5.96 & 212 & 14 & C* \\
I16562-3959-C1 & 2 & 254.92327 & -40.06196 & 0.43 & 0.24 & 592 & 329 & -144 & -14.56 & 4.32 & 436 & 69 & C \\
I17016-4124-C1 & 1 & 256.29549 & -41.48524 & 1.49 & 0.94 & 4702 & 2958 & 101 & - & - & - & - & A \\
I17016-4124-C3 & 2 & 256.29666 & -41.48537 & 0.74 & 0.36 & 2351 & 1138 & 163 & -27.99 & 4.00 & 302 & 33 & D \\
I17016-4124-C2 & 3 & 256.29625 & -41.48415 & 0.22 & 0.17 & 683 & 531 & 176 & - & - & - & - & A \\
I17143-3700-C1 & 1 & 259.43941 & -37.05362 & 0.50 & 0.38 & 5902 & 4497 & 133 & -32.57 & 4.37 & 258 & 22 & C* \\
I17143-3700-C1 & 2 & 259.43929 & -37.05328 & 0.86 & 0.53 & 10117 & 6183 & -174 & - & - & - & - & A \\
I17160-3707-C1 & 1 & 259.86426 & -37.18547 & 0.48 & 0.38 & 4934 & 3948 & 153 & -67.98 & 7.28 & 647 & 175 & C \\
I17160-3707-C2 & 2 & 259.86252 & -37.18150 & 0.36 & 0.26 & 3701 & 2714 & -178 & -68.90 & 10.00 & 272 & 27 & A \\
I17175-3544-C3 & 1 & 260.22251 & -35.78431 & 0.91 & 0.50 & 1204 & 665 & 62 & -7.35 & 2.15 & 159 & 9 & D \\
I17175-3544-C2 & 2 & 260.22151 & -35.78307 & 0.79 & 0.74 & 1045 & 982 & 129 & - & - & - & - & C \\
I17175-3544-C1 & 3 & 260.22254 & -35.78292 & 1.18 & 0.58 & 1552 & 760 & 67 & - & - & - & - & B \\
I17175-3544-C2 & 4 & 260.22168 & -35.78278 & 0.48 & 0.31 & 634 & 412 & 121 & - & - & - & - & C \\
I17175-3544-C1 & 5 & 260.22249 & -35.78258 & 0.89 & 0.46 & 1172 & 602 & 105 & - & - & - & - & B \\
I17175-3544-C1 & 6 & 260.22275 & -35.78257 & 0.41 & 0.14 & 539 & 190 & 66 & - & - & - & - & B \\
I17175-3544-C1 & 7 & 260.22252 & -35.78226 & 0.19 & 0.17 & 253 & 222 & 120 & -6.21 & 10.00 & 798 & 255 & B \\
I17204-3636-C1 & 1 & 260.95920 & -36.64989 & 0.50 & 0.31 & 1648 & 1020 & -166 & -17.65 & 9.47 & 234 & 26 & A \\
I17220-3609-C1 & 1 & 261.35518 & -36.21275 & 0.34 & 0.29 & 2705 & 2318 & 119 & - & - & - & - & D \\
I17220-3609-C1 & 2 & 261.35543 & -36.21274 & 0.26 & 0.19 & 2125 & 1546 & 116 & - & - & - & - & A \\
I17220-3609-C1 & 3 & 261.35526 & -36.21232 & 0.91 & 0.50 & 7342 & 4057 & 69 & - & - & - & - & A \\
I17220-3609-C1 & 4 & 261.35564 & -36.21228 & 0.41 & 0.31 & 3284 & 2512 & -178 & - & - & - & - & B \\
I17220-3609-C1 & 5 & 261.35509 & -36.21204 & 0.22 & 0.17 & 1739 & 1352 & -177 & -92.88 & 4.94 & 177 & 12 & C \\
I17271-3439\_1-C2 & 1 & 262.61076 & -34.69651 & 0.26 & 0.22 & 966 & 791 & 156 & -13.09 & 8.86 & 198 & 27 & C \\
I17271-3439\_1-C1 & 2 & 262.60917 & -34.69607 & 0.41 & 0.24 & 1493 & 878 & -154 & -19.28 & 2.84 & 103 & 6 & D \\
I17545-2357-C1 & 1 & 269.39499 & -23.96824 & 0.55 & 0.29 & 1656 & 864 & 68 & 7.24 & 2.37 & 125 & 10 & D \\
I17599-2148-C1 & 1 & 270.75307 & -21.80283 & 0.65 & 0.48 & 2229 & 1651 & -178 & - & - & - & - & A \\
I18032-2032-C3 & 1 & 271.56222 & -20.52871 & 0.50 & 0.36 & 2419 & 1728 & 94 & 4.41 & 3.96 & 175 & 11 & D \\
I18032-2032-C2 & 2 & 271.56199 & -20.52762 & 1.10 & 0.62 & 5299 & 2995 & 99 & - & - & - & - & A \\
I18032-2032-C4 & 3 & 271.56169 & -20.52700 & 0.82 & 0.50 & 3917 & 2419 & -153 & - & - & - & - & A \\
I18032-2032-C1 & 4 & 271.56121 & -20.52575 & 0.17 & 0.17 & 806 & 806 & 103 & 3.64 & 6.48 & 189 & 15 & C \\
I18032-2032-C1 & 5 & 271.56114 & -20.52543 & 0.72 & 0.67 & 3456 & 3226 & 88 & - & - & - & - & A \\
I18056-1952-C1 & 1 & 272.15941 & -19.86408 & 0.82 & 0.48 & 6887 & 4051 & 125 & - & - & - & - & D \\
I18056-1952-C1 & 2 & 272.15923 & -19.86397 & 0.70 & 0.43 & 5874 & 3646 & 131 & - & - & - & - & D \\
I18110-1854-C1 & 1 & 273.50369 & -18.89062 & 0.36 & 0.26 & 1141 & 837 & 160 & - & - & - & - & C \\
I18116-1646-C1 & 1 & 273.64941 & -16.76006 & 0.19 & 0.14 & 741 & 556 & 161 & 47.46 & 9.68 & 102 & 16 & C \\
I18116-1646-C1 & 2 & 273.64933 & -16.75996 & 0.24 & 0.22 & 926 & 834 & 77 & 48.46 & 5.67 & 198 & 25 & A \\
I18434-0242-C1 & 1 & 281.51575 & -2.65621 & 1.46 & 0.96 & 6969 & 4570 & -172 & - & - & - & - & A \\
I18469-0132-C1 & 1 & 282.38772 & -1.48432 & 1.03 & 0.74 & 5067 & 3653 & 78 & 87.26 & 10.00 & 722 & 211 & B* \\
I18479-0005-C1 & 1 & 282.62806 & -0.03312 & 0.67 & 0.62 & 8649 & 8031 & 55 & - & - & - & - & C \\
I18507+0110-C1 & 1 & 283.32779 & 1.24862 & 0.46 & 0.43 & 1473 & 1395 & 47 & - & - & - & - & A \\
I18507+0110-C1 & 2 & 283.32793 & 1.24886 & 0.50 & 0.26 & 1628 & 853 & 161 & - & - & - & - & A \\
I18507+0110-C1 & 3 & 283.32817 & 1.24892 & 0.17 & 0.14 & 543 & 465 & -164 & 56.06 & 2.75 & 157 & 7 & C \\
I18507+0110-C1 & 4 & 283.32795 & 1.24935 & 0.24 & 0.19 & 775 & 620 & -173 & - & - & - & - & A \\
I18507+0110-C1 & 5 & 283.32733 & 1.24943 & 0.43 & 0.31 & 1395 & 1008 & 132 & - & - & - & - & D \\
I18507+0110-C1 & 6 & 283.32785 & 1.24947 & 0.36 & 0.29 & 1163 & 930 & 124 & - & - & - & - & C \\
I18507+0110-C1 & 7 & 283.32760 & 1.24954 & 0.58 & 0.17 & 1860 & 543 & 140 & - & - & - & - & D \\
I18507+0110-C1 & 8 & 283.32764 & 1.24981 & 0.17 & 0.14 & 543 & 465 & -164 & 59.76 & 7.59 & 528 & 80 & D \\
I19078+0901-C1 & 1 & 287.55586 & 9.10297 & 0.72 & 0.48 & 8273 & 5515 & 93 & 3.49 & 5.12 & 138 & 9 & A \\
I19078+0901-C3 & 2 & 287.55292 & 9.10314 & 0.53 & 0.46 & 6067 & 5239 & 158 & 12.94 & 8.17 & 222 & 15 & C \\
I19078+0901-C3 & 3 & 287.55361 & 9.10319 & 0.26 & 0.14 & 3033 & 1655 & 117 & 13.78 & 8.26 & 199 & 17 & D \\
I19078+0901-C2 & 4 & 287.55486 & 9.10368 & 1.80 & 1.08 & 20682 & 12409 & 107 & 17.54 & 7.90 & 452 & 125 & D \\
I19078+0901-C3 & 5 & 287.55365 & 9.10333 & 0.24 & 0.12 & 2758 & 1379 & 133 & 12.15 & 10.00 & 236 & 25 & D \\
I19078+0901-C1 & 6 & 287.55597 & 9.10351 & 0.74 & 0.60 & 8549 & 6894 & 171 & 2.14 & 7.92 & 307 & 54 & D \\
I19078+0901-C2 & 7 & 287.55534 & 9.10353 & 0.65 & 0.50 & 7446 & 5791 & 178 & 8.56 & 6.02 & 348 & 64 & A \\
I19078+0901-C5 & 8 & 287.55554 & 9.10449 & 0.91 & 0.60 & 10479 & 6894 & -162 & 3.80 & 5.09 & 161 & 9 & C \\
I19078+0901-C4 & 9 & 287.55472 & 9.10521 & 0.34 & 0.24 & 3861 & 2758 & -176 & 6.89 & 5.11 & 182 & 15 & D \\
I19095+0930-C1 & 1 & 287.97498 & 9.59732 & 0.89 & 0.82 & 8054 & 7401 & 154 & 41.45 & 8.28 & 508 & 116 & C* \\
I19097+0847-C2 & 1 & 288.03759 & 8.87062 & 0.67 & 0.38 & 5262 & 3007 & 53 & 59.34 & 7.98 & 142 & 18 & A \\
I19097+0847-C1 & 2 & 288.03840 & 8.87082 & 0.53 & 0.48 & 4134 & 3758 & 139 & 59.39 & 7.39 & 350 & 41 & A \\
\enddata
\tablenotetext{$a$}{ “A” denotes \textcolor{black}{HMFs} that are associated with jet-like outflow. “B” denotes \textcolor{black}{HMFs} that are associated with wide angle outflow. “C” denotes hot cores without outflow or with weak outflow. “D” denotes shell-like \textcolor{black}{HMFs}, which appear more like shells around \HII\ regions and lack significant associated 1.3 mm continuum emission. “*” denotes that \ch3cn{} emission spatially coincide with H30$\alpha$ emission.}

\end{deluxetable*}

\section{\HII\ region catalog}
\restartappendixnumbering
Using \astrodendro , the \HII\ regions were identified in the $\mathrm{H}30\alpha$ emission maps. 
 The \astrodendro\ parameters were configured with 
$\mathtt{min\_delta} = 3\,\sigma$ and $\mathtt{min\_value} = 5\,\sigma$. Table \ref{tab:hii} column (3)-(8) are H30$\alpha$ line integrated  intensity peak coordinates RA, DEC, the intensity-weighted second moments along the major and minor axes, position angle from measured from north to east, and size ($=\sqrt{\mathrm{Mag\_FWHM}\times\mathrm{Min\_FWHM}}$). 
Additionally, Gaussian fittings were performed to derive the physical parameters including central velocity (column (9)), line width (column (10)), and amplitude (column (11)) from the 
averaged $\mathrm{H}30\alpha$ spectra of these identified \HII\ region. Figure \ref{fig:HIIcla} presents the relation between their size and linewidth.

\begin{figure}[!htb]
    \centering
    \includegraphics[width=9cm]{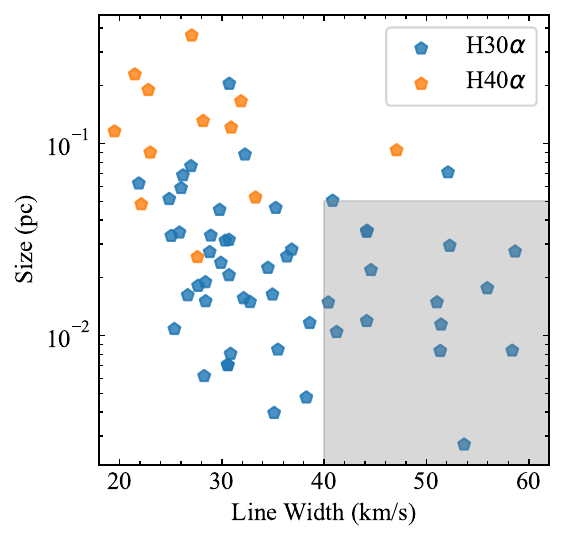}
    \caption{Relation between the size and the line width of the ionized gas structure. The sizes and line widths of these ionized gas structures were measured from H30$\alpha$ (blue pentagons) or H40$\alpha$ (orange pentagons). Those located in gray shadow are regard as candidate HC\,H\,{\sc ii} region cadi, while the rest are regard as candidate UC\,H\,{\sc ii} regions.
    }
    \label{fig:HIIcla}
\end{figure}


\startlongtable
\tabletypesize{\scriptsize}

\begin{deluxetable*}{lrlrllrrrrlr}
\renewcommand{\arraystretch}{0.8}
\tablecaption{Physical parameters of the \HII\ regions \label{tab:hii}}
\tablehead{
\colhead{Field} & \colhead{ID} & \colhead{ra} & \colhead{dec} & \colhead{Maj\_FWHM} & 
\colhead{Min\_FWHM} & \colhead{PA} & \colhead{size} & \colhead{Vlsr} & 
\colhead{Linewidth} & \colhead{Amplitude}& \colhead{Type} \\[-7pt]
\colhead{-} & \colhead{-} & \colhead{(\degree)} & \colhead{(\degree)} & \colhead{(\arcsec)} & 
\colhead{(\arcsec)} & \colhead{(\degree)} & \colhead{(AU)} & \colhead{(\kms)} & 
\colhead{(\kms)} & \colhead{(Jy\,beam$^{-1}$)}& \colhead{-}  
}
\decimalcolnumbers
\startdata
I12320-6122 & 1 & 188.72186 & -61.66111 & 0.58 & 0.55 & 102 & 2402 & -39.25 & 38.59 & 0.1611 & UC \\
I12326-6245 & 1 & 188.89618 & -63.04219 & 0.43 & 0.36 & -149 & 1718 & -70.70 & 51.35 & 2.0680 & HC \\
I12572-6316\_1 & 1 & 195.10028 & -63.54215 & 0.34 & 0.26 & 106 & 3349 & 24.60 & 26.66 & 0.0348 & UC \\
I13471-6120 & 1 & 207.67410 & -61.58617 & 1.10 & 0.96 & 77 & 5335 & -57.62 & 36.36 & 0.2147 & UC \\
I15254-5621 & 1 & 232.33076 & -56.52285 & 0.72 & 0.70 & 137 & 3076 & -69.48 & 51.03 & 0.4050 & HC \\
I15290-5546 & 1 & 233.22192 & -55.93608 & 3.31 & 1.58 & 175 & 18103 & -88.52 & 32.26 & 0.0605 & UC \\
I15290-5546 & 2 & 233.22174 & -55.93468 & 1.97 & 1.58 & 172 & 14101 & -89.90 & 26.19 & 0.0610 & UC \\
I15411-5352 & 1 & 236.24718 & -54.03887 & 4.73 & 1.68 & -169 & 6825 & -44.91 & 25.05 & 0.0325 & UC \\
I15439-5449 & 1 & 236.95870 & -54.97589 & 2.26 & 1.56 & 120 & 6446 & -53.21 & 30.34 & 0.0157 & UC \\
I15502-5302 & 1 & 238.52668 & -53.19468 & 2.23 & 1.61 & -146 & 10409 & -93.86 & 40.83 & 0.2007 & UC \\
I15520-5234 & 1 & 238.95153 & -52.71856 & 2.02 & 0.72 & -166 & 3072 & -41.96 & 40.41 & 0.0860 & HC \\
I15520-5234 & 2 & 238.95222 & -52.71853 & 0.43 & 0.36 & 137 & 983 & -40.89 & 38.26 & 0.2972 & UC \\
I15520-5234 & 3 & 238.95273 & -52.71836 & 0.72 & 0.55 & 149 & 1659 & -35.95 & 30.85 & 0.1597 & UC \\
I15567-5236 & 1 & 240.13813 & -52.74651 & 2.62 & 0.60 & 172 & 6515 & -107.51 & 30.70 & 0.0472 & UC \\
I15567-5236 & 2 & 240.13701 & -52.74624 & 1.51 & 0.43 & 66 & 4260 & -112.36 & 30.70 & 0.0492 & UC \\
I15567-5236 & 3 & 240.13788 & -52.74604 & 2.18 & 0.89 & -137 & 7266 & -113.18 & 44.19 & 0.0734 & HC \\
I16037-5223 & 1 & 241.90610 & -52.51699 & 1.63 & 1.22 & 165 & 12801 & -80.06 & 21.88 & 0.0495 & UC \\
I16037-5223 & 2 & 241.91185 & -52.51662 & 1.10 & 0.98 & -150 & 9329 & -78.59 & 29.78 & 0.0335 & UC \\
I16037-5223 & 3 & 241.90884 & -52.51493 & 1.15 & 0.98 & -176 & 9546 & -78.77 & 35.27 & 0.1029 & UC \\
I16060-5146 & 1 & 242.46825 & -51.91604 & 0.82 & 0.58 & 124 & 3751 & -86.72 & 27.68 & 0.1173 & UC \\
I16060-5146 & 2 & 242.46944 & -51.91539 & 0.65 & 0.55 & 116 & 3234 & -92.48 & 32.13 & 0.8439 & UC \\
I16060-5146 & 3 & 242.46843 & -51.91499 & 0.94 & 0.72 & 61 & 4528 & -83.13 & 44.58 & 0.2740 & HC \\
I16060-5146 & 4 & 242.46947 & -51.91501 & 0.48 & 0.41 & 103 & 2458 & -89.82 & 44.16 & 1.7127 & HC \\
I16065-5158 & 1 & 242.58458 & -52.10186 & 9.62 & 6.53 & 97 & 34294 & - & - & - & UC \\
I16071-5142 & 1 & 242.74752 & -51.83989 & 3.53 & 3.50 & -142 & 18536 & - & - & - & UC \\
I16164-5046 & 1 & 245.04616 & -50.88731 & 0.84 & 0.82 & -168 & 3377 & -64.26 & 34.95 & 0.5454 & UC \\
I16164-5046 & 2 & 245.04541 & -50.88729 & 0.41 & 0.34 & 178 & 1447 & -56.29 & 30.59 & 0.2154 & UC \\
I16172-5028\_1 & 1 & 245.26199 & -50.58711 & 0.24 & 0.19 & -141 & 816 & -50.54 & 35.11 & 0.1552 & UC \\
I16172-5028\_1 & 2 & 245.26227 & -50.58686 & 0.48 & 0.43 & 112 & 1724 & -65.45 & 58.39 & 0.9693 & HC \\
I16172-5028\_1 & 3 & 245.26265 & -50.58671 & 0.48 & 0.24 & 105 & 1270 & -49.75 & 28.27 & 0.3915 & UC \\
I16313-4729 & 1 & 248.72676 & -47.59372 & 0.46 & 0.41 & 140 & 2156 & -75.86 & 41.22 & 0.0880 & HC \\
I16348-4654 & 1 & 249.62350 & -47.00986 & 2.30 & 2.11 & -176 & 25017 & - & - & - & UC \\
I16351-4722 & 1 & 249.71030 & -47.46750 & 0.72 & 0.55 & 178 & 1454 & -32.97 & 30.56 & 0.0544 & UC \\
I16445-4459 & 1 & 252.02158 & -45.08617 & 3.26 & 2.90 & 143 & 23901 & - & - & - & UC \\
I16458-4512 & 1 & 252.37521 & -45.29569 & 0.84 & 0.67 & -154 & 3117 & -53.05 & 28.41 & 0.0575 & UC \\
I16562-3959 & 1 & 254.92345 & -40.06212 & 0.43 & 0.41 & 132 & 559 & -12.35 & 53.69 & 0.1069 & HC \\
I17016-4124 & 1 & 256.29668 & -41.48531 & 0.58 & 0.53 & 119 & 1744 & -32.31 & 35.48 & 0.1706 & UC \\
I17143-3700 & 1 & 259.43941 & -37.05362 & 0.55 & 0.41 & 161 & 5621 & -34.22 & 28.81 & 0.0952 & UC \\
I17160-3707 & 1 & 259.86464 & -37.18394 & 7.94 & 6.79 & 125 & 75496 & - & - & - & UC \\
I17175-3544 & 1 & 260.22276 & -35.78394 & 4.42 & 3.62 & -169 & 5291 & - & - & - & UC \\
I17204-3636 & 1 & 260.95967 & -36.64992 & 3.14 & 2.93 & -157 & 9967 & - & - & - & UC \\
I17220-3609 & 1 & 261.35514 & -36.21268 & 1.42 & 1.25 & 156 & 10626 & -94.13 & 24.85 & 0.1792 & UC \\
I17271-3439\_1 & 1 & 262.60909 & -34.69599 & 1.68 & 0.96 & 165 & 4656 & -12.53 & 34.50 & 0.0460 & UC \\
I17545-2357 & 1 & 269.39508 & -23.96822 & 0.79 & 0.70 & 171 & 2232 & 7.80 & 25.36 & 0.0474 & UC \\
I17599-2148 & 1 & 270.75571 & -21.80328 & 10.94 & 5.64 & 135 & 27080 & - & - & - & UC \\
I17599-2148 & 2 & 270.75320 & -21.80239 & 3.38 & 2.93 & 57 & 10815 & - & - & - & UC \\
I18032-2032 & 1 & 271.56224 & -20.52864 & 0.91 & 0.72 & 138 & 3917 & 10.67 & 28.39 & 0.0481 & UC \\
I18056-1952 & 1 & 272.15920 & -19.86394 & 2.45 & 2.06 & -171 & 19041 & - & - & - & UC \\
I18110-1854 & 1 & 273.50417 & -18.89025 & 2.81 & 1.66 & 142 & 6847 & 41.31 & 28.92 & 0.0215 & UC \\
I18116-1646 & 1 & 273.64796 & -16.76078 & 16.44 & 9.14 & -162 & 47339 & - & - & - & UC \\
I18434-0242 & 1 & 281.51637 & -2.65608 & 3.98 & 1.61 & 62 & 12109 & 100.62 & 26.01 & 0.0472 & UC \\
I18469-0132 & 1 & 282.38772 & -1.48431 & 0.55 & 0.46 & 162 & 2357 & 96.12 & 51.43 & 0.0722 & HC \\
I18479-0005 & 1 & 282.62771 & -0.03361 & 0.62 & 0.48 & 166 & 7104 & 16.37 & 25.85 & 0.1962 & UC \\
I18479-0005 & 2 & 282.62793 & -0.03349 & 0.26 & 0.22 & 127 & 3089 & 12.60 & 32.75 & 0.1048 & UC \\
I18479-0005 & 3 & 282.62783 & -0.03328 & 0.46 & 0.34 & 134 & 4942 & 2.52 & 29.90 & 0.0966 & UC \\
I18479-0005 & 4 & 282.62797 & -0.03283 & 1.44 & 1.03 & -171 & 15753 & 12.18 & 26.98 & 0.0331 & UC \\
I18479-0005 & 5 & 282.62912 & -0.03243 & 5.30 & 2.04 & 51 & 42317 & 19.33 & 30.72 & 0.0179 & UC \\
I18507+0110 & 1 & 283.32742 & 1.24949 & 1.25 & 1.01 & 88 & 3643 & 50.51 & 55.95 & 1.0739 & HC \\
I19078+0901 & 1 & 287.55592 & 9.10361 & 2.14 & 0.74 & -151 & 14615 & 14.16 & 52.11 & 0.5309 & UC \\
I19078+0901 & 2 & 287.55371 & 9.10325 & 0.55 & 0.50 & 99 & 6067 & 14.71 & 52.28 & 0.4158 & HC \\
I19078+0901 & 3 & 287.55479 & 9.10353 & 0.74 & 0.50 & 137 & 7170 & 15.66 & 44.16 & 0.8211 & HC \\
I19078+0901 & 4 & 287.55478 & 9.10521 & 0.50 & 0.48 & -173 & 5791 & -3.09 & 36.83 & 0.3094 & UC \\
I19095+0930 & 1 & 287.97498 & 9.59728 & 0.67 & 0.60 & 171 & 5660 & 66.82 & 58.67 & 0.1135 & HC \\
I19097+0847 & 1 & 288.03641 & 8.86911 & 5.71 & 4.37 & 131 & 39275 & - & - & - & UC \\
\enddata
\tablecomments{“-” indicates that H30$\alpha$ is not detected in ALMA band 6. Therefore, we use H40$\alpha$ (ALMA band 3 ) line integrated intensity for \HII\ region identification.}
\end{deluxetable*}

\section{images for all selected fields}
\restartappendixnumbering

In this section, we present the complete set of images for our sample. Each field is displayed in a two-panel format with the following configuration: 
Left panel, the ATOMS Band 3 3.0\,mm continuum (gray contours) and H40$\alpha$ line integrated intensity (green color); 
Right panel, the QUARKS Band 6 1.3\,mm continuum (gray contours) and H30$\alpha$ line integrated intensity (green color). Additionally, the QUARKS \CO\ outflow (red and blue colors) and \ch3cn\ (12$_3$--11$_3$) line integrated intensity (orange contours) are overlaid on both panels. The details can be found in the caption of Figure \ref{fig:sample_image}. 

\begin{figure*}
\gridline{
    \fig{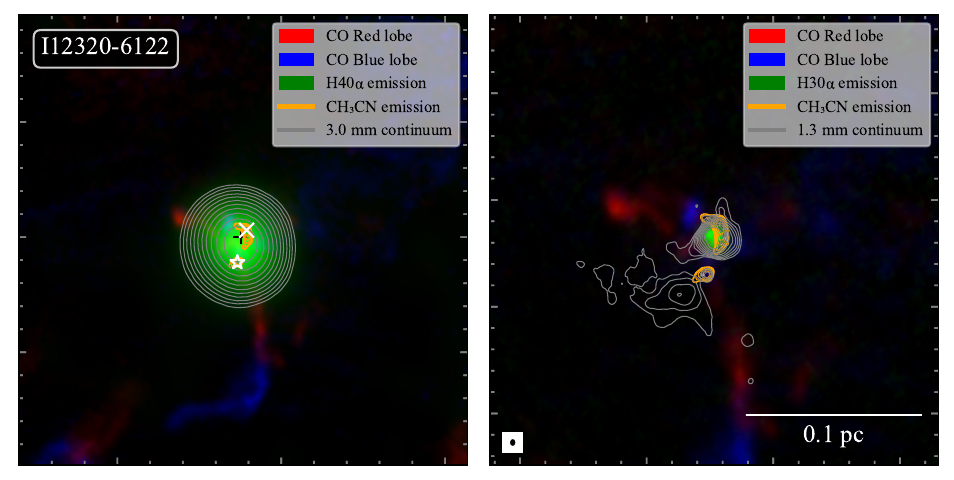}{0.49\textwidth}{}
    \fig{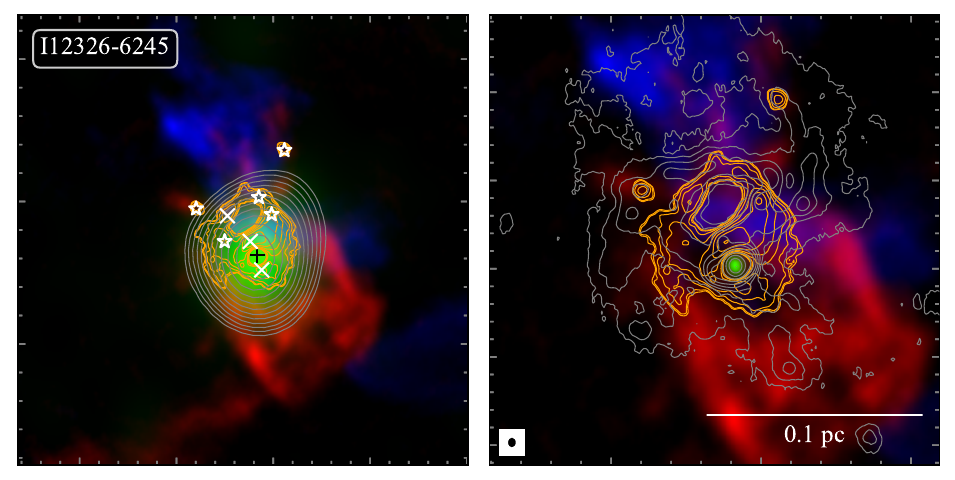}{0.49\textwidth}{}
}
\vspace{-7mm}

\gridline{
    \fig{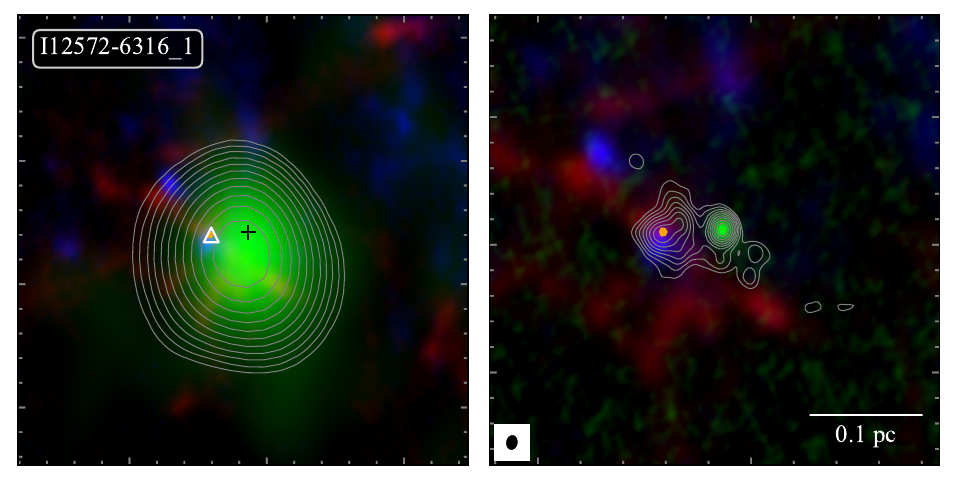}{0.49\textwidth}{}
    \fig{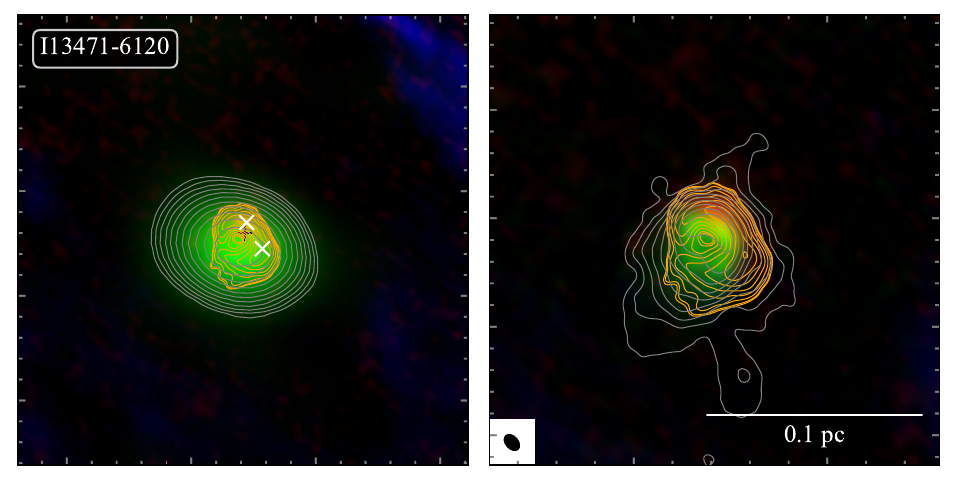}{0.49\textwidth}{}
}
\vspace{-7mm}

\gridline{
    \fig{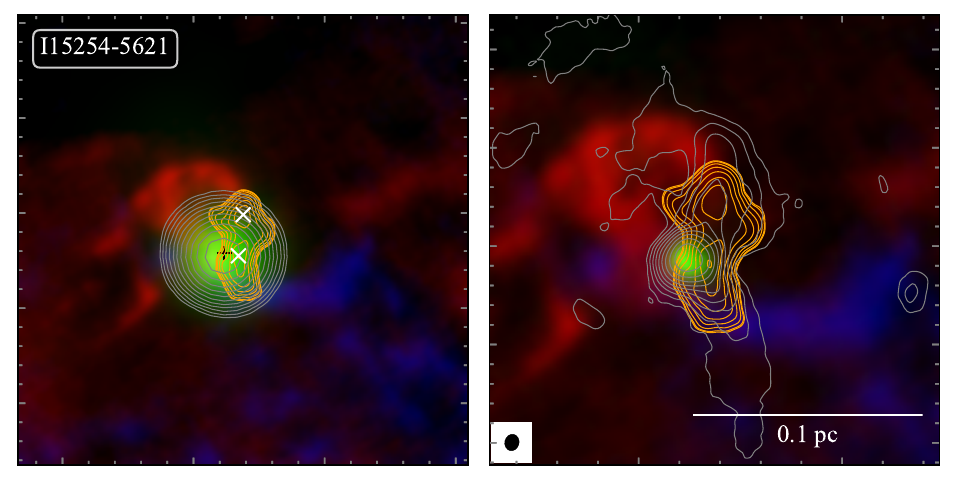}{0.49\textwidth}{}
    \fig{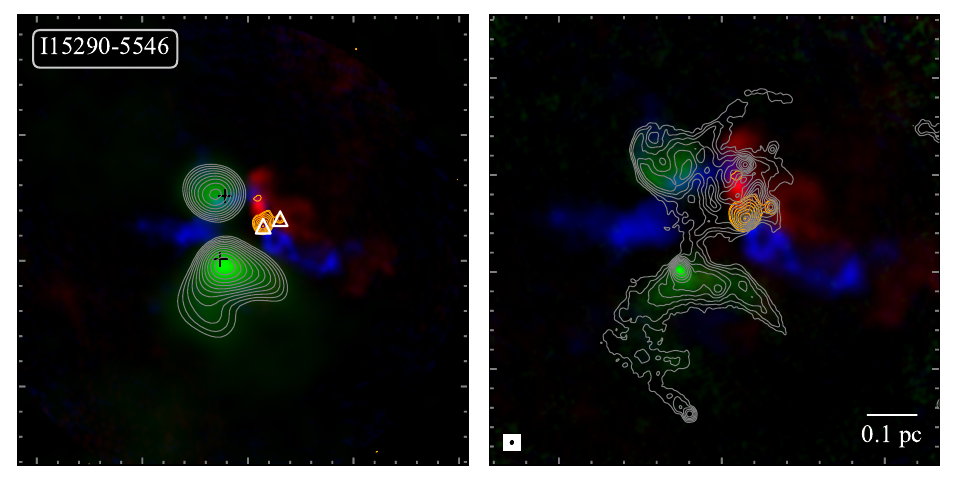}{0.49\textwidth}{}
}
\vspace{-7mm}

\gridline{
    \fig{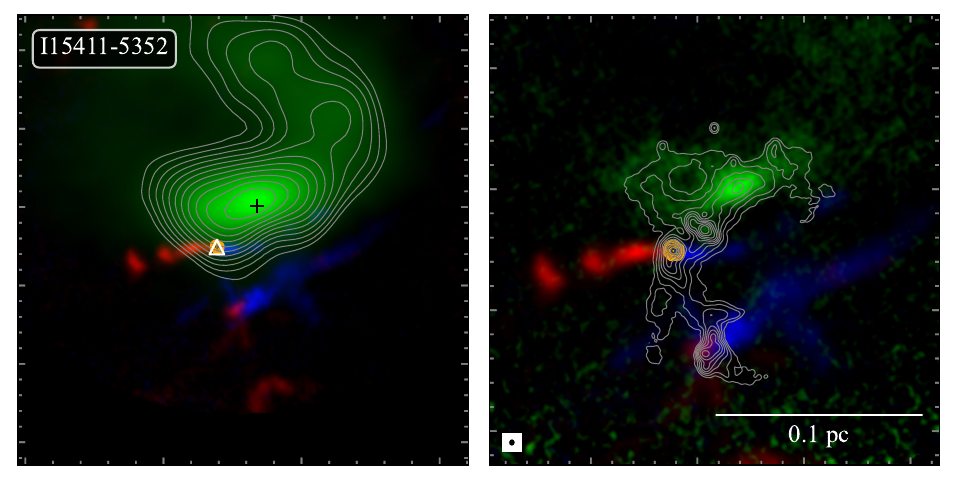}{0.49\textwidth}{}
    \fig{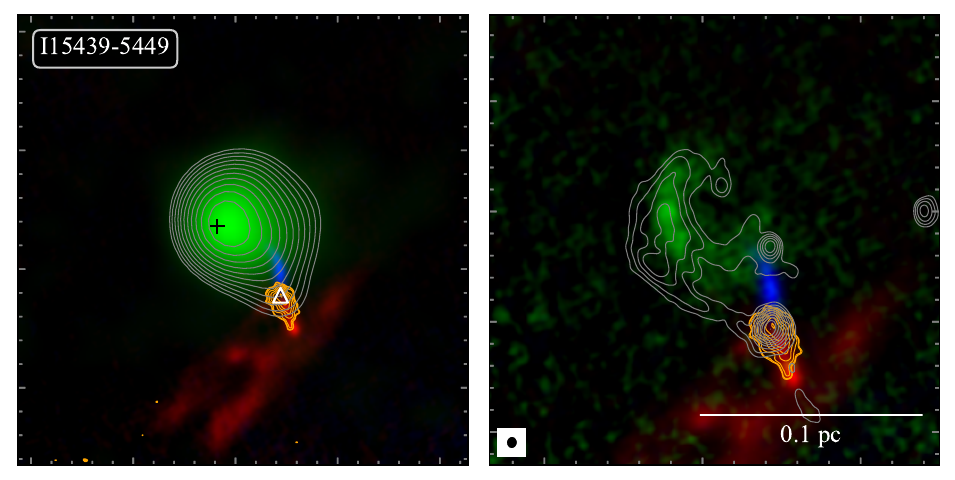}{0.49\textwidth}{}
}
\vspace{-7mm}

\gridline{
    \fig{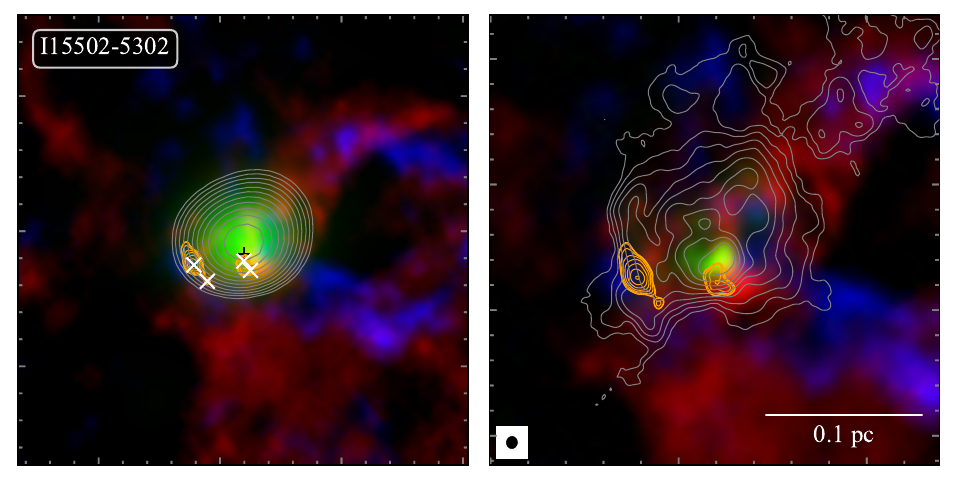}{0.49\textwidth}{}
    \fig{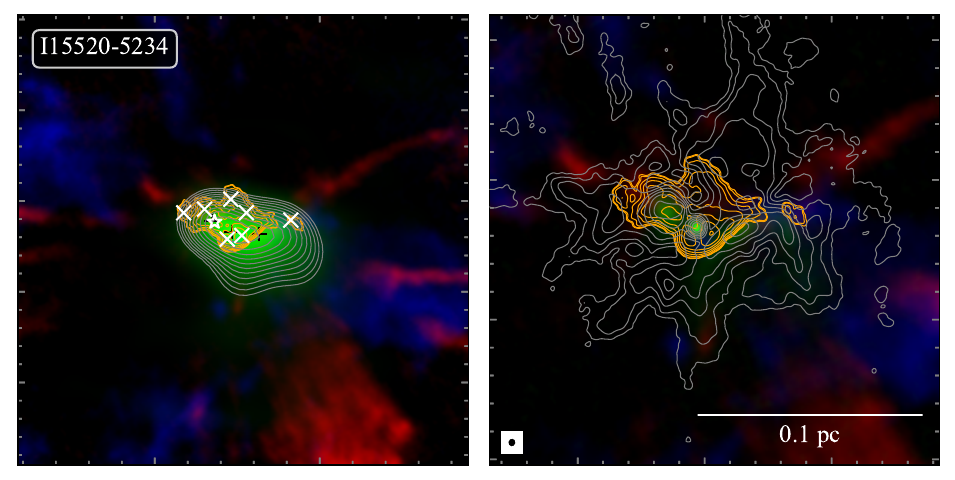}{0.49\textwidth}{}
}
\vspace{-7mm}
\caption{\it -- continued}
\label{fig:sample_image}
\end{figure*}

\addtocounter{figure}{-1}
\begin{figure*}
\gridline{
    \fig{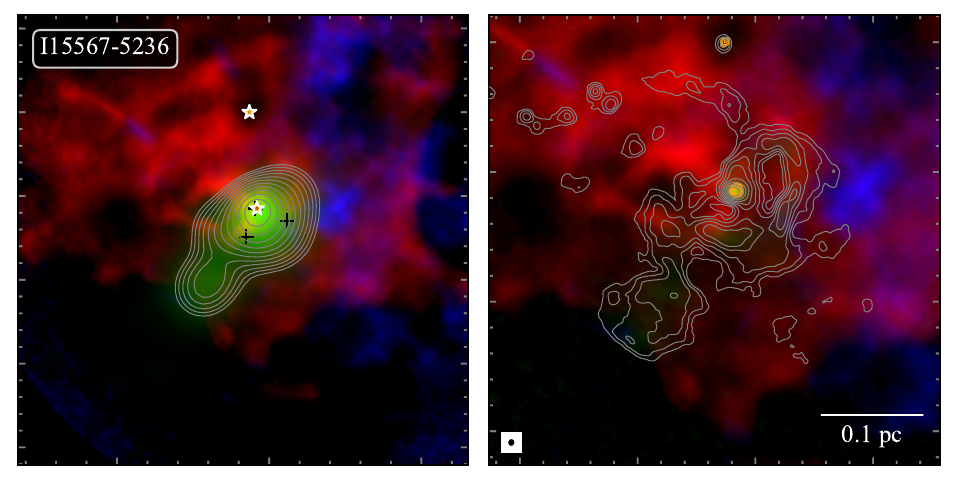}{0.49\textwidth}{}
    \fig{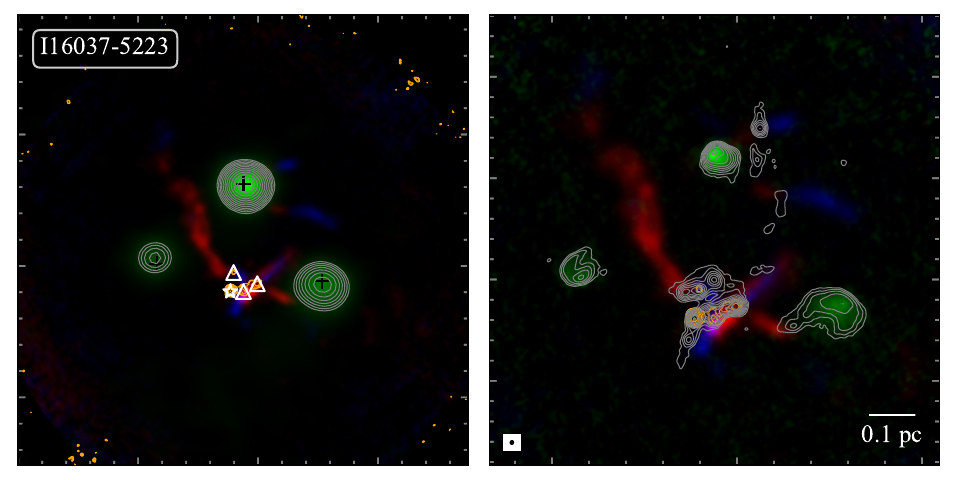}{0.49\textwidth}{}
}
\vspace{-7mm}

\gridline{
    \fig{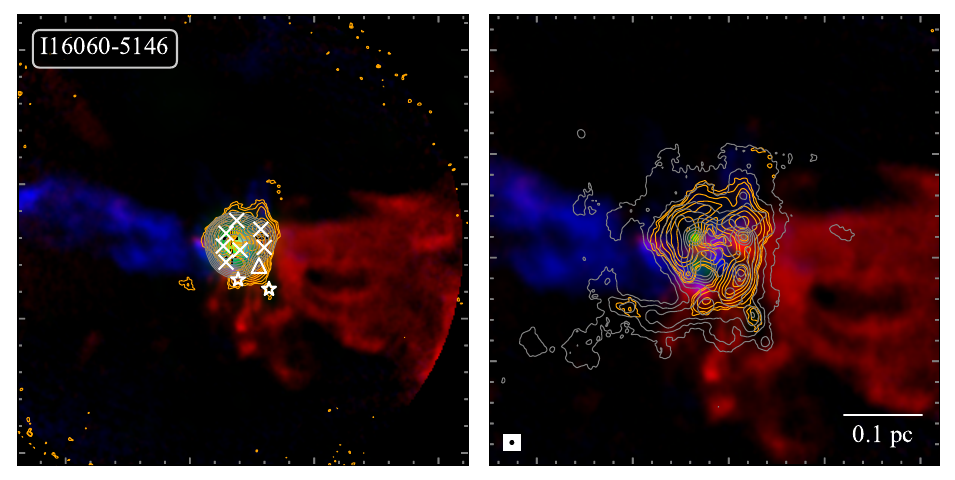}{0.49\textwidth}{}
    \fig{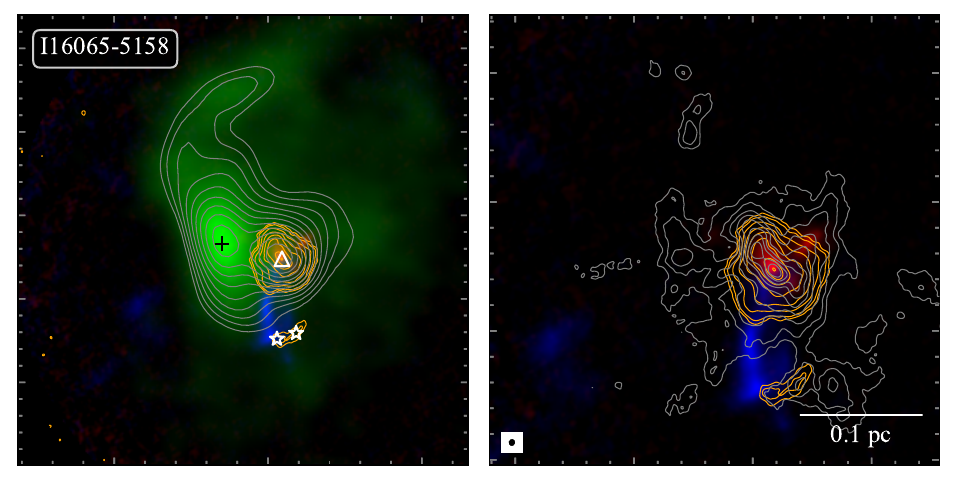}{0.49\textwidth}{}
}
\vspace{-7mm}

\gridline{
    \fig{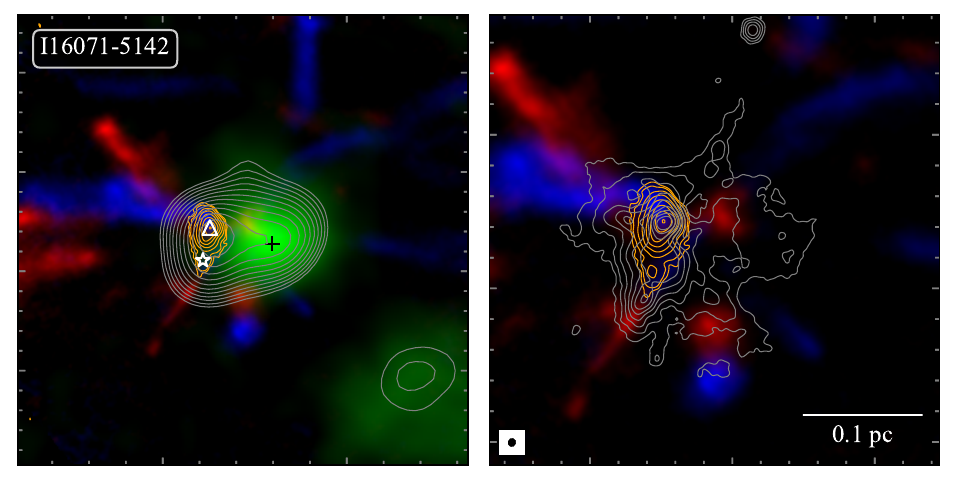}{0.49\textwidth}{}
    \fig{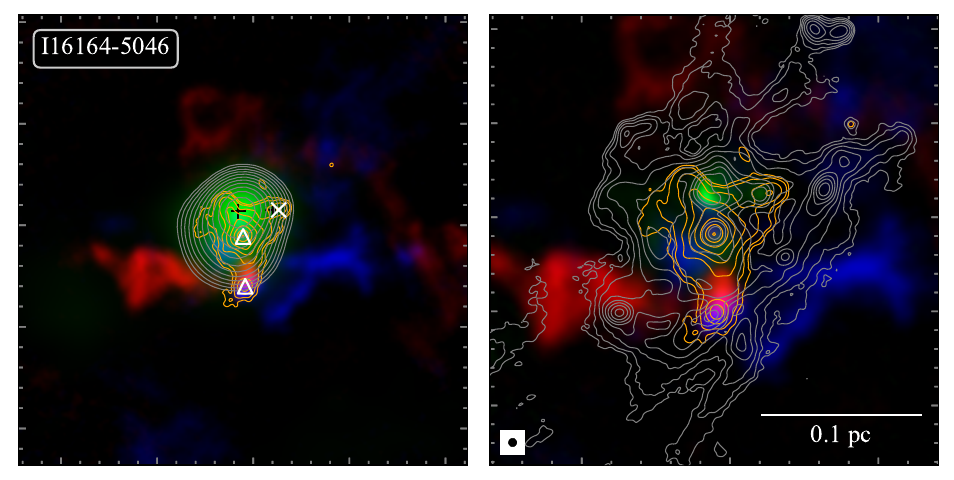}{0.49\textwidth}{}
}
\vspace{-7mm}

\gridline{
    \fig{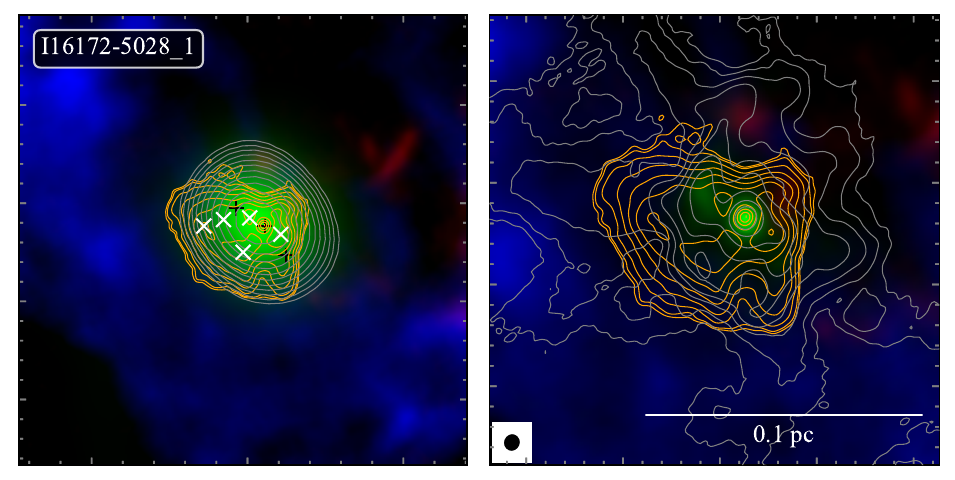}{0.49\textwidth}{}
    \fig{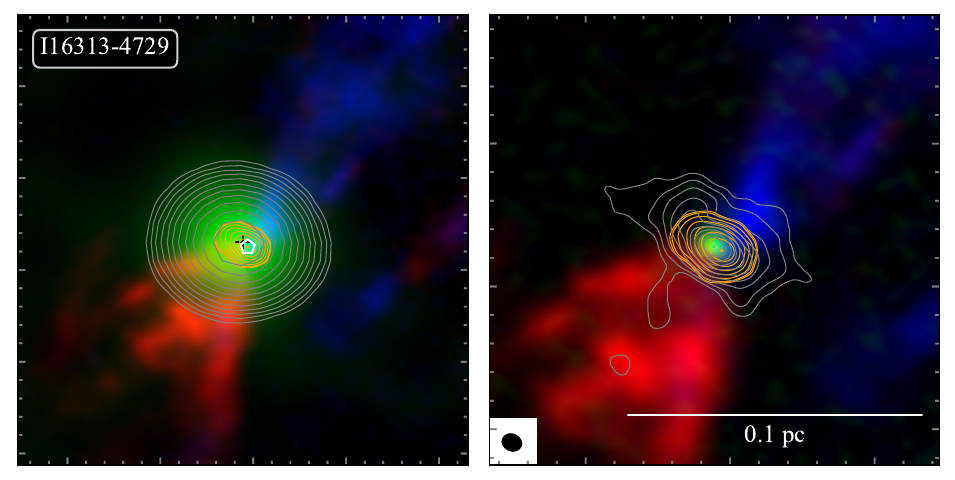}{0.49\textwidth}{}
}
\vspace{-7mm}

\gridline{
    \fig{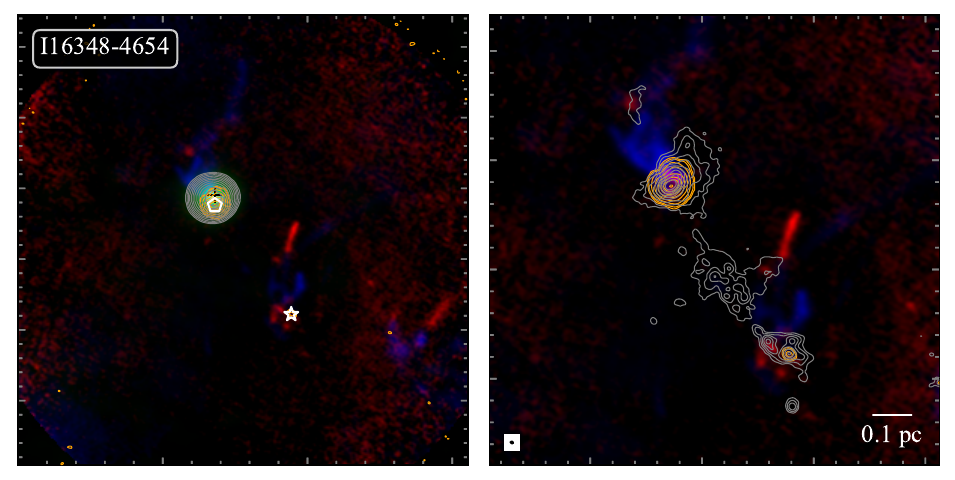}{0.49\textwidth}{}
    \fig{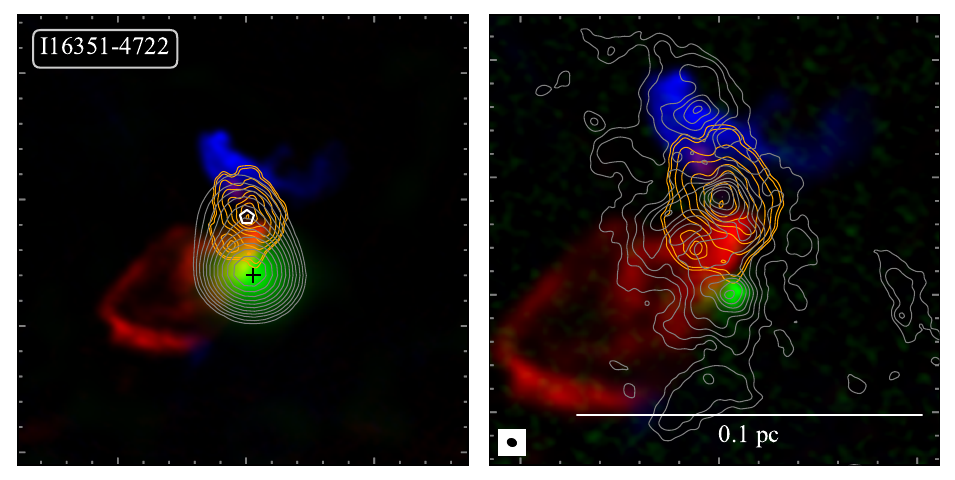}{0.49\textwidth}{}
}
\vspace{-7mm}
\caption{\it -- continued}
\end{figure*}

\addtocounter{figure}{-1}
\begin{figure*}
\gridline{
    \fig{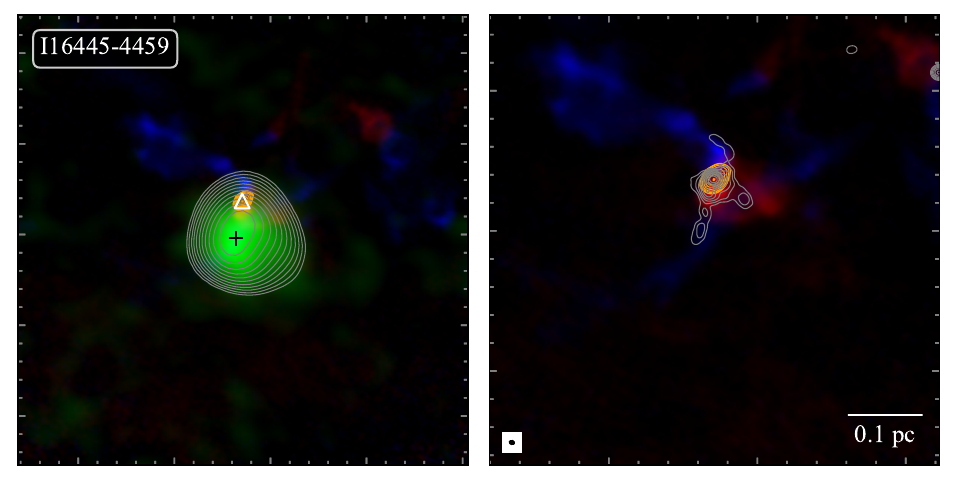}{0.49\textwidth}{}
    \fig{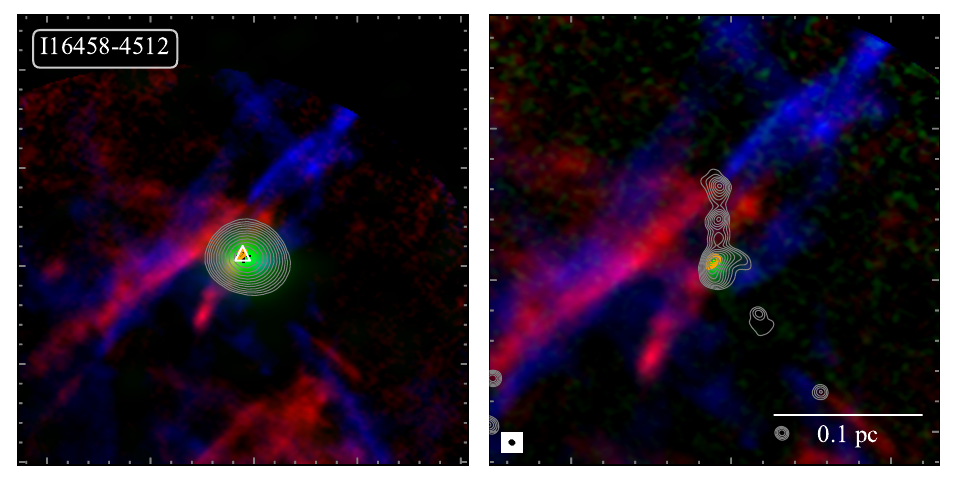}{0.49\textwidth}{}
}
\vspace{-7mm}

\gridline{
    \fig{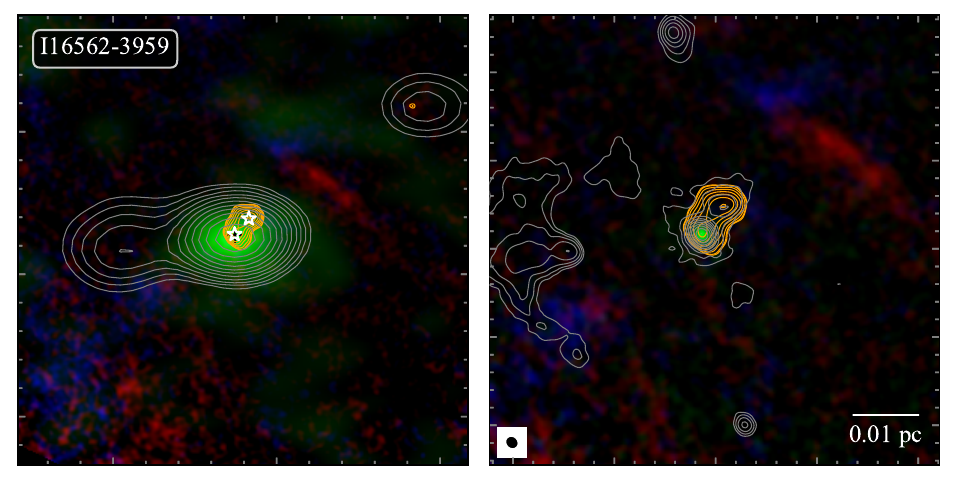}{0.49\textwidth}{}
    \fig{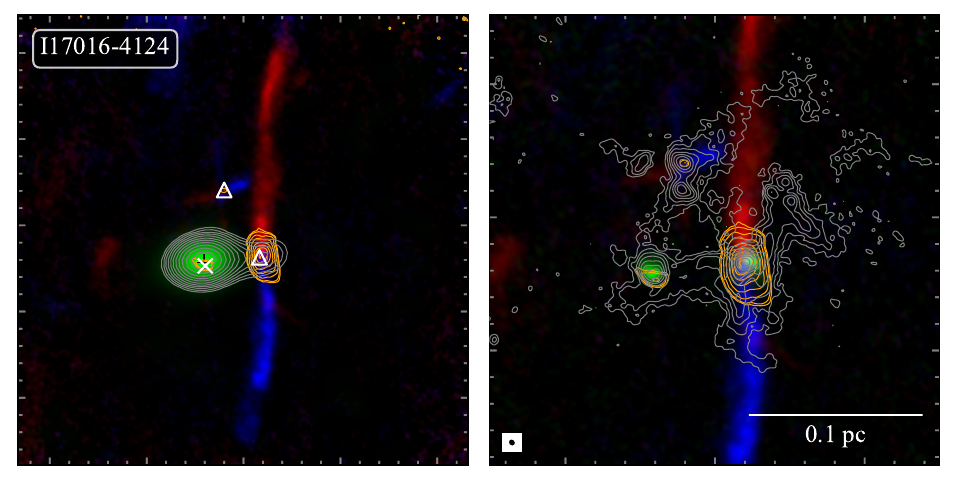}{0.49\textwidth}{}
}
\vspace{-7mm}

\gridline{
    \fig{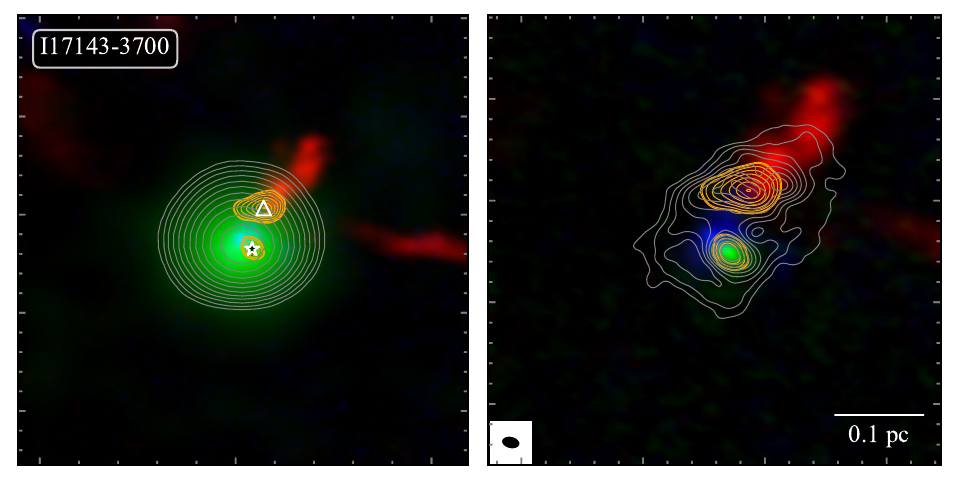}{0.49\textwidth}{}
    \fig{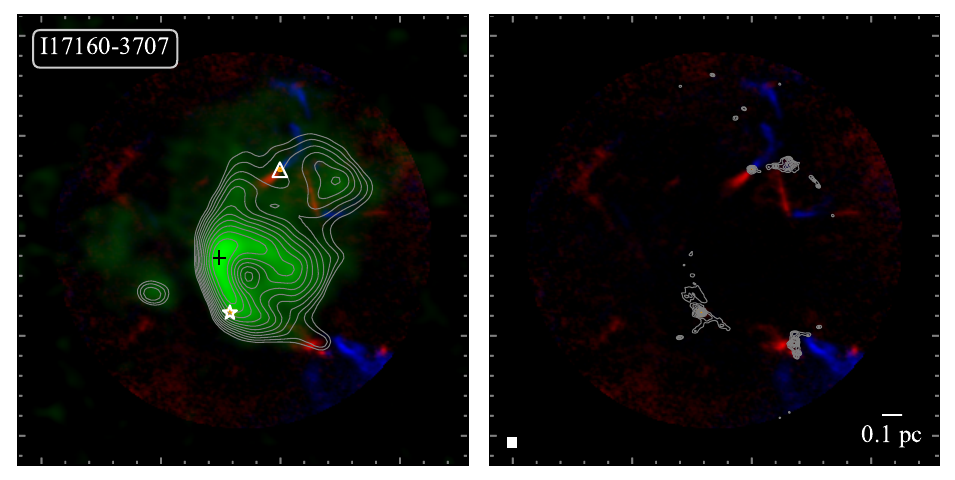}{0.49\textwidth}{}
}
\vspace{-7mm}

\gridline{
    \fig{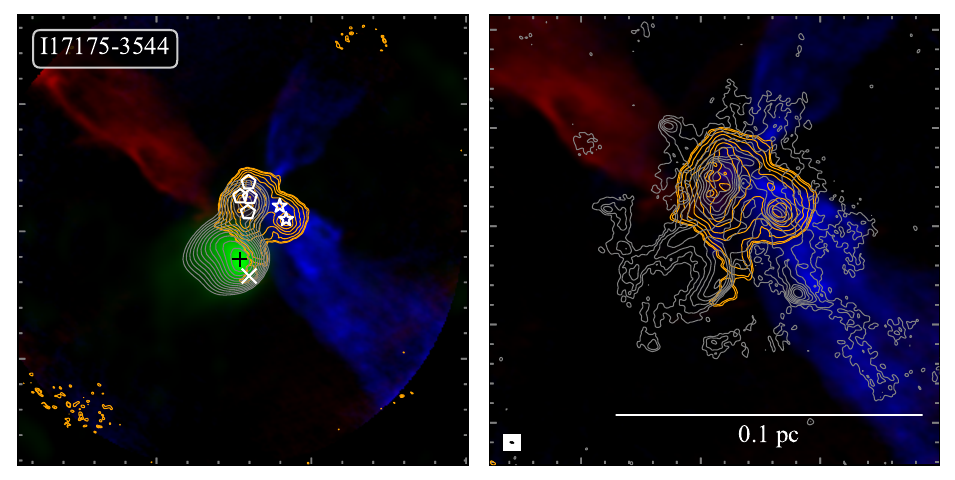}{0.49\textwidth}{}
    \fig{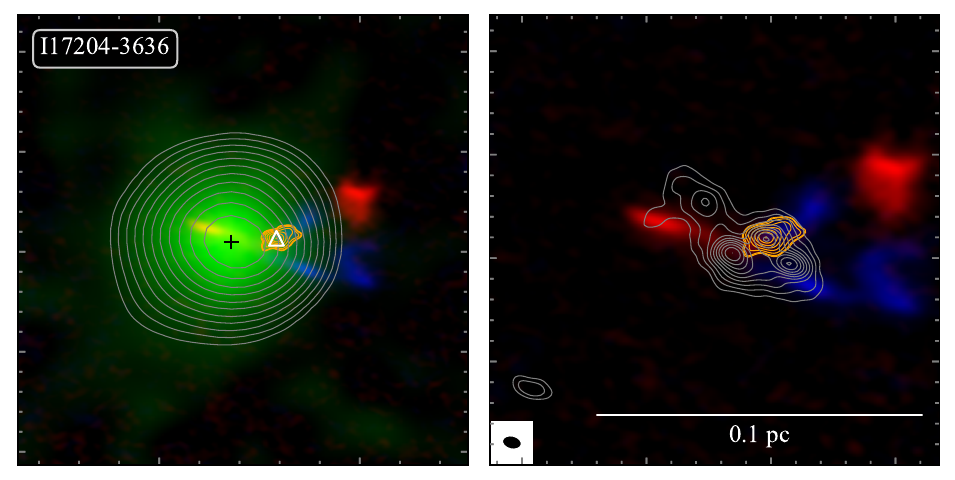}{0.49\textwidth}{}
}
\vspace{-7mm}

\gridline{
    \fig{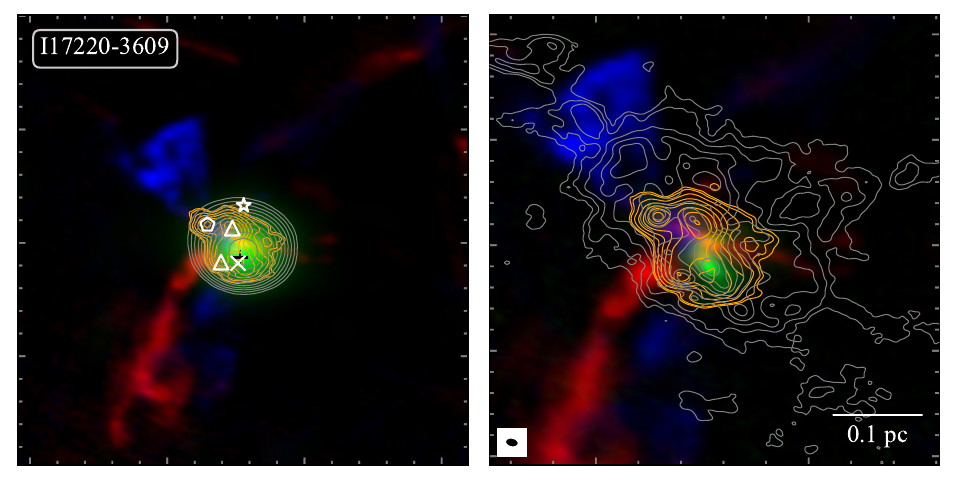}{0.49\textwidth}{}
    \fig{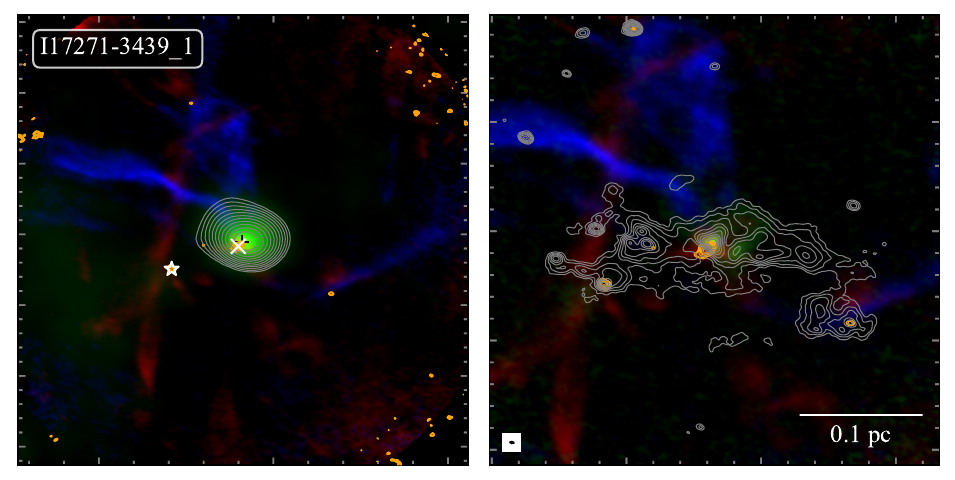}{0.49\textwidth}{}
}
\vspace{-7mm}
\caption{\it -- continued}
\end{figure*}

\addtocounter{figure}{-1}
\begin{figure*}
\gridline{
    \fig{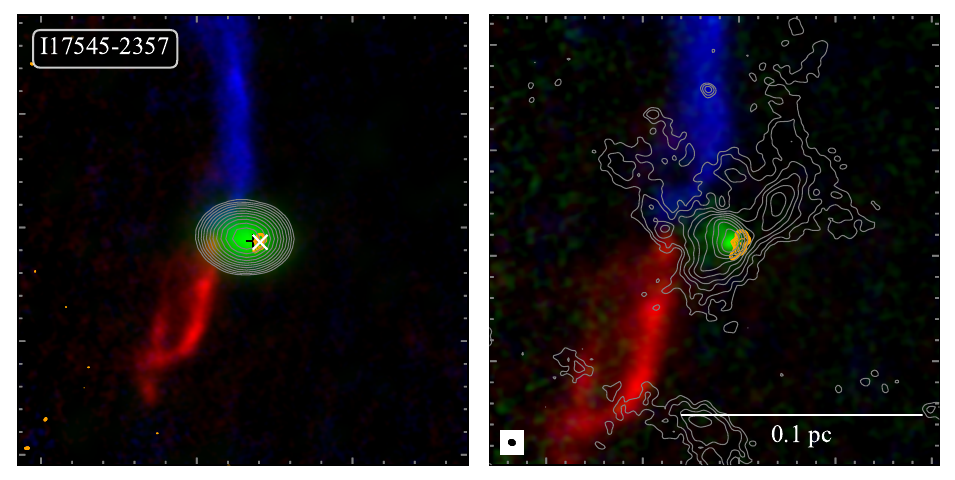}{0.49\textwidth}{}
    \fig{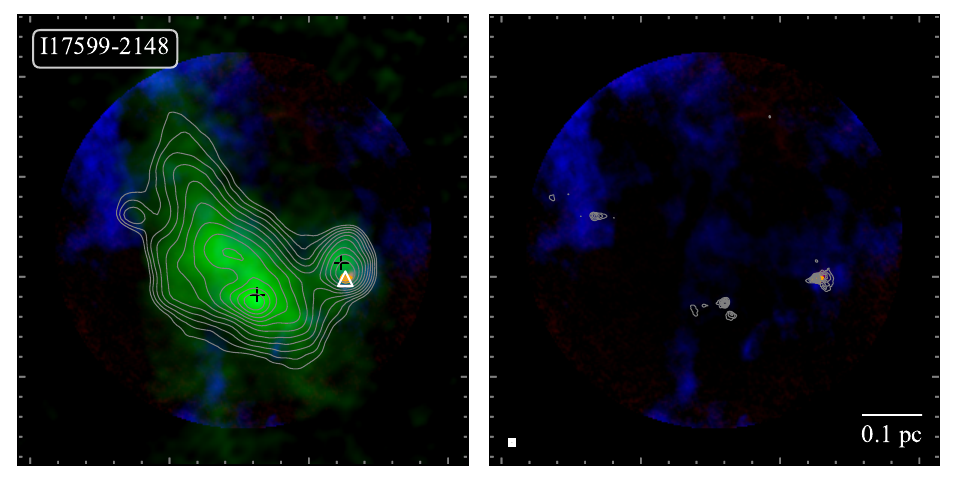}{0.49\textwidth}{}
}
\vspace{-7mm}

\gridline{
    \fig{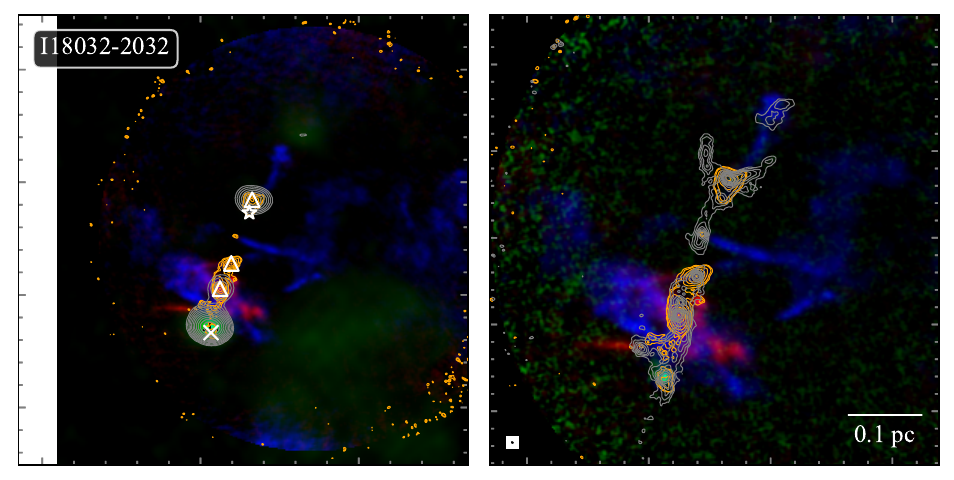}{0.49\textwidth}{}
    \fig{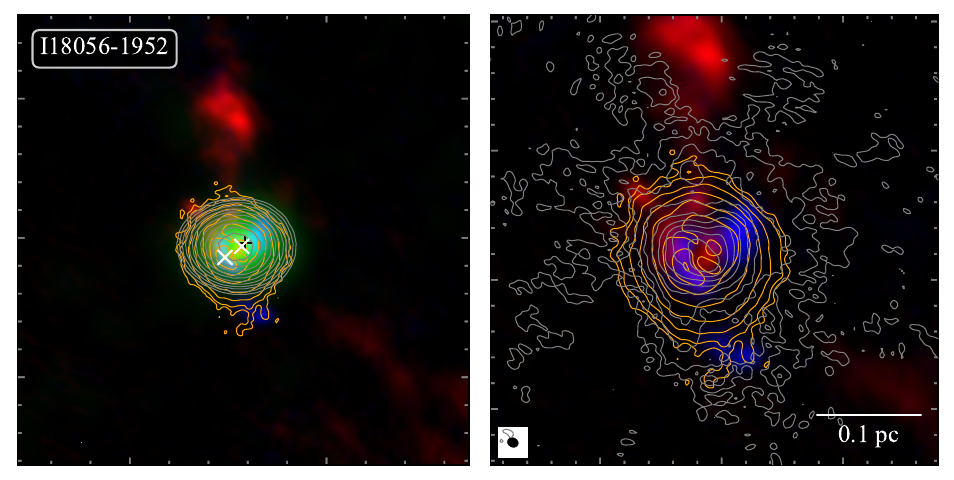}{0.49\textwidth}{}
}
\vspace{-7mm}

\gridline{
    \fig{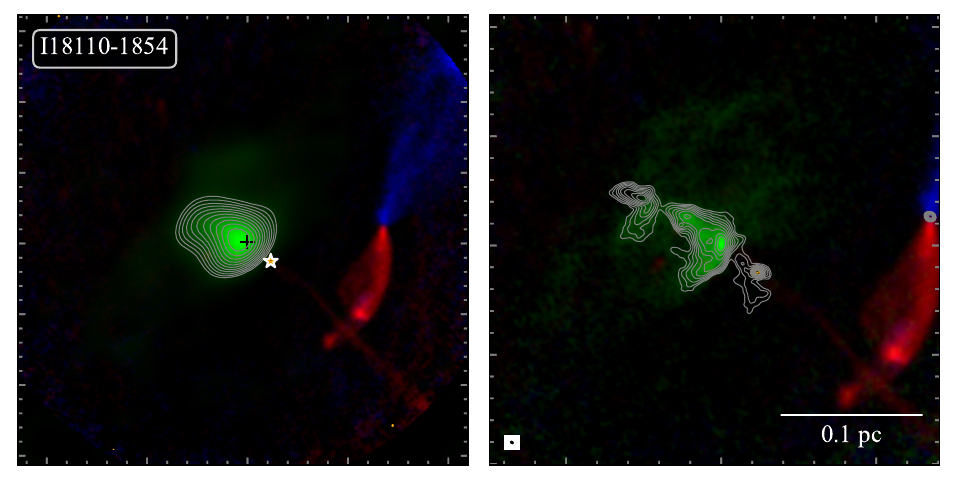}{0.49\textwidth}{}
    \fig{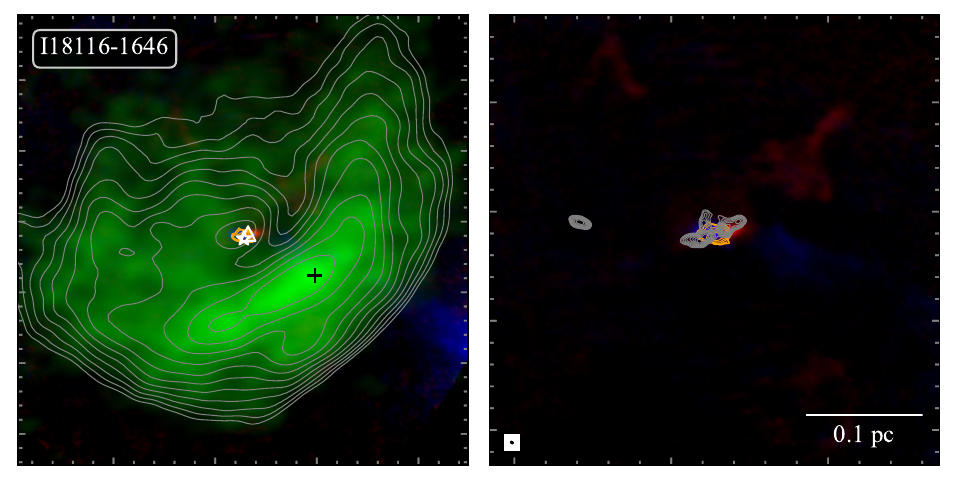}{0.49\textwidth}{}
}
\vspace{-7mm}

\gridline{
    \fig{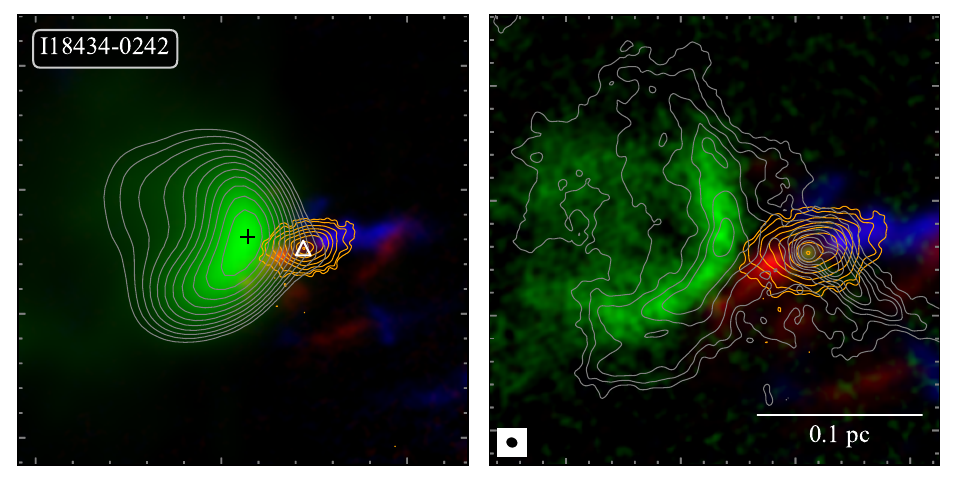}{0.49\textwidth}{}
    \fig{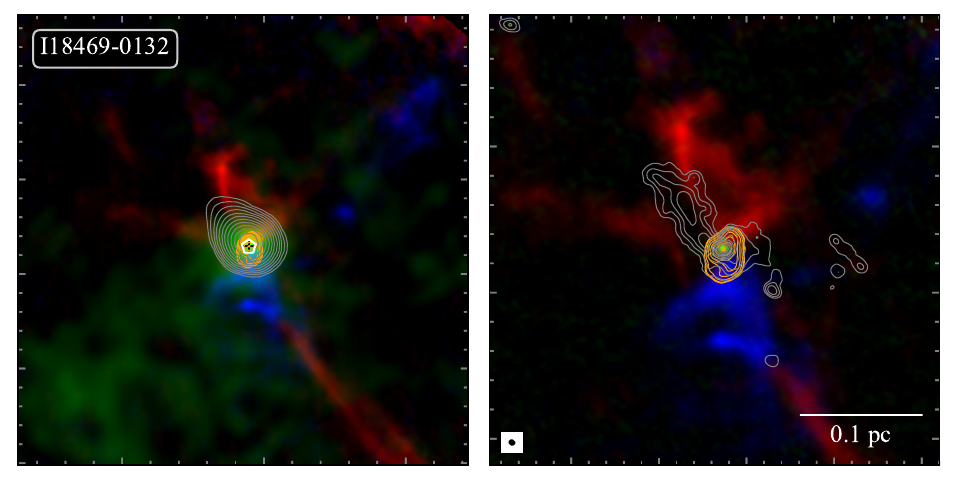}{0.49\textwidth}{}
}
\vspace{-7mm}

\gridline{
    \fig{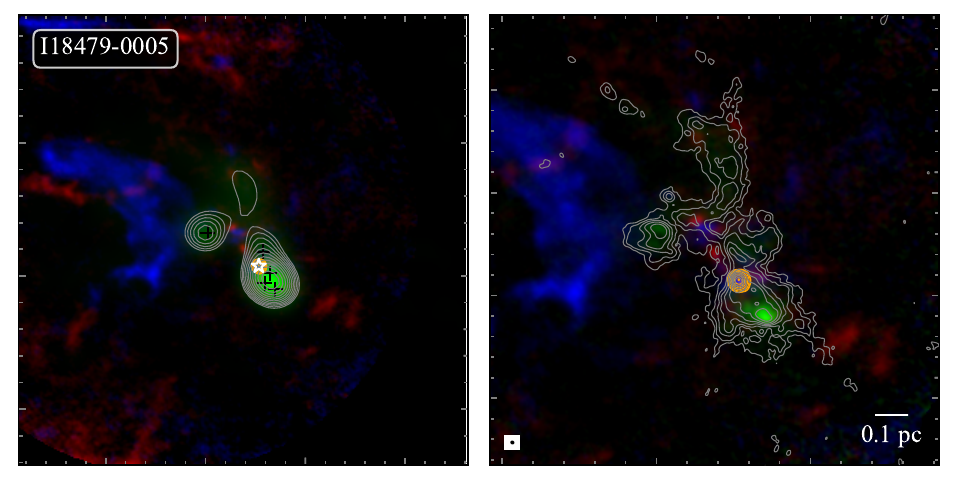}{0.49\textwidth}{}
    \fig{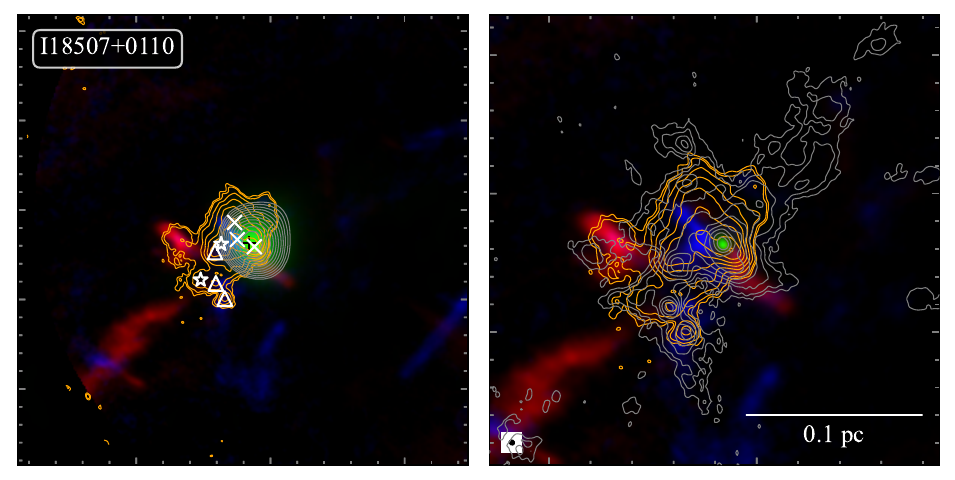}{0.49\textwidth}{}
}
\vspace{-7mm}
\caption{\it -- continued}
\label{fig:hotcores_outflows}
\end{figure*}

\addtocounter{figure}{-1}
\begin{figure*}
\gridline{
    \fig{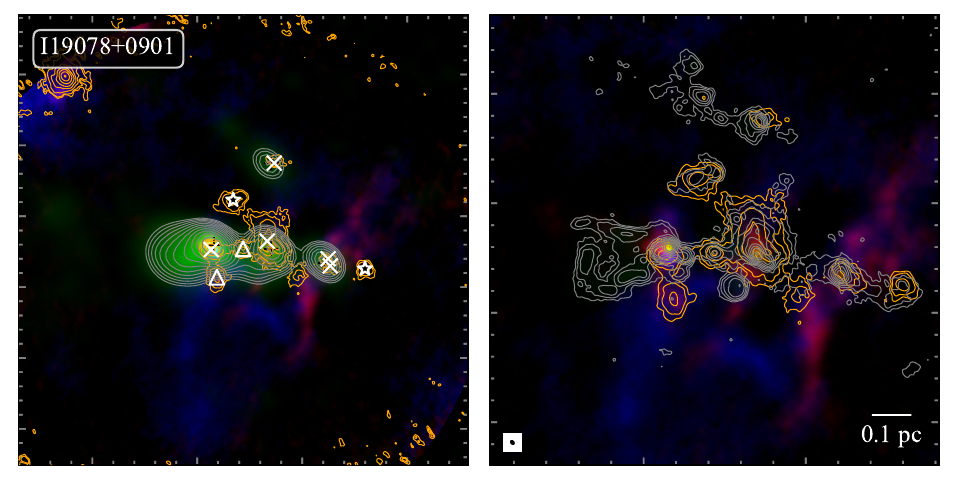}{0.49\textwidth}{}
    \fig{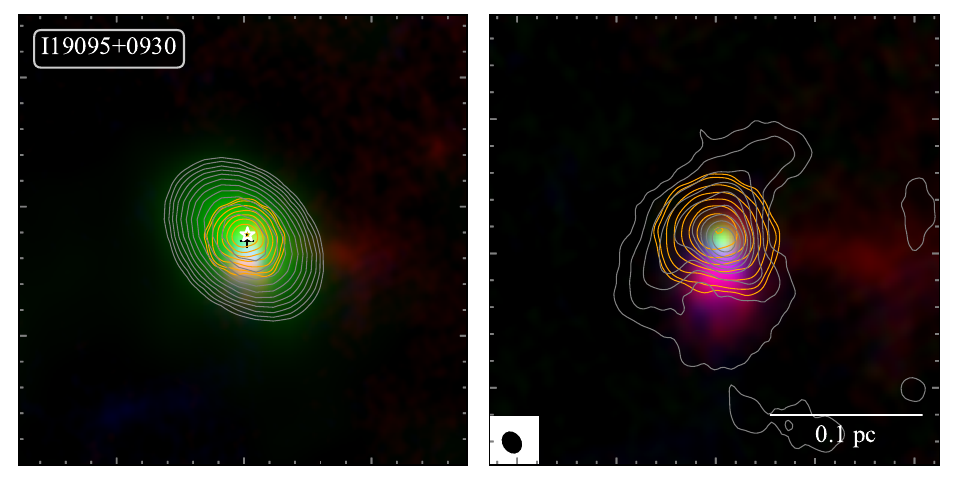}{0.49\textwidth}{}
}
\vspace{-7mm}

\gridline{
    \fig{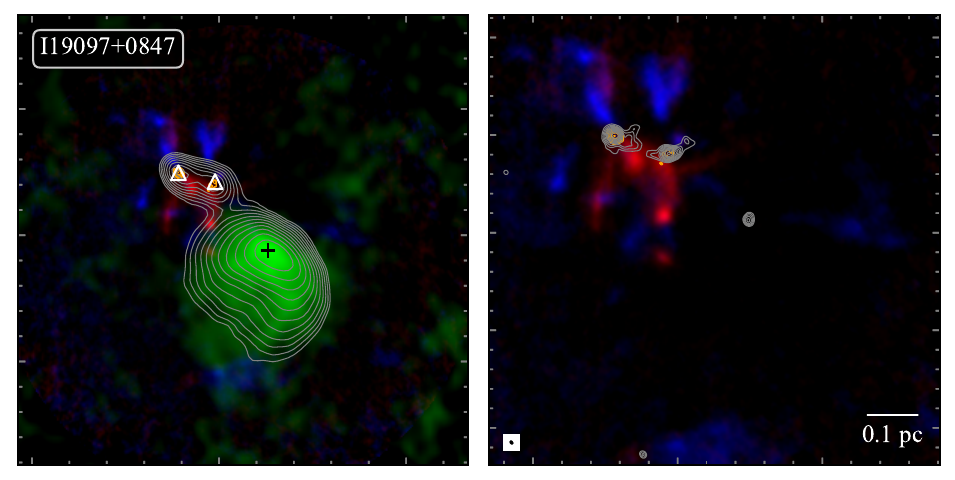}{0.49\textwidth}{}
    \hspace{0.5\textwidth}
}
\vspace{-7mm}
\caption{Images of the hydrogen  recombination line emssion and the outflow for our sample. In the left panels for each source, the background shows the three-color image composed by \CO\ outflow red lobe (red), H40$\alpha$ (green) and \CO\ outflow blue lobe (blue), and the gray contours and orange contours represent the 3 mm continuum and the \ch3cn\,(12$_3$--11$_3$) integrated line emission, respectively. In the right panels for each source, the green background and gray contours represent H30$\alpha$ and 1.3 mm continuum, and the others are the same as the panel on the left.
H30$\alpha$/H40$\alpha$ integrated velocity range is [Vlsr-40\,\kms, Vlsr+40\,\kms], where Vlsr is central velocity. For \ch3cn\,(12$_3$--11$_3$), the integrated velocity ranges are different for different fields. Similarly, distinct integrated velocity ranges were adopted for \CO\ to avoid contamination and clearly reveal the outflow features.
The contour levels were plotted from $3\, \sigma$  to the peak intensity of the field, with 8 logarithmically spaced contours between these values. White markers mark the positions of the \textcolor{black}{HMFs} (jet-like outflow: triangle; wide-angle outflow: pentagon; no/weak outflow: star; shell-like: cross), while black plus signs indicate the locations of the \HII\ regions. The synthesized beams are shown in the lower left corner of the right panel, while the scale bar is indicated in the lower right corner.}
\end{figure*}

\end{document}